\documentclass[11pt,a4paper]{article}
\pdfoutput=1
\usepackage{jcappub}

\usepackage{graphicx}
\usepackage{caption}
\usepackage{color}
\usepackage{amsmath,amsfonts,amsthm}

\newcommand{\change}[1]{{\color{black} #1}}
\newcommand{\newchange}[1]{{\color{black} #1}}
\newcommand{\beq}{\begin{equation}}
\newcommand{\eeq}{\end{equation}}
\newcommand{\ben}{\begin{eqnarray}}
\newcommand{\een}{\end{eqnarray}}
\newcommand{\bi}{\begin{itemize}}
\newcommand{\ei}{\end{itemize}}
\newcommand{\nn}{\nonumber}

\newcommand{\ie}{\textit{i.e.}}
\newcommand{\eg}{\textit{e.g.}}
\newcommand{\etc}{\textit{etc}}

\newcommand{\citeeq}[1]{Eq.~(\ref{#1})}

\newcommand{\citesec}[1]{Sec.~\ref{#1}}

\newcommand{\citeapp}[1]{App.~\ref{#1}}

\newcommand{\citetab}[1]{Tab.~\ref{#1}}
\newcommand{\citefig}[1]{Fig.~\ref{#1}}

\graphicspath{{./NewFigs/}{./}}

\subheader{LUPM:18-022}

\title{Anatomy of Eddington-like inversion methods in the context of dark matter searches}
\author{Thomas Lacroix,}
\author{Martin Stref,}
\author{and Julien Lavalle}

\affiliation{Laboratoire Univers et Particules de Montpellier (LUPM),\\
Universit\'e de Montpellier \& CNRS, Place Eug\`ene Bataillon, 34095 Montpellier Cedex 05, France}

\emailAdd{thomas.lacroix@umontpellier.fr}
\emailAdd{martin.stref@umontpellier.fr}
\emailAdd{lavalle@in2p3.fr}

\abstract{
Irrespective of the dark matter (DM) candidate, several potentially observable signatures derive
from the velocity distribution of DM in halos, in particular in the Milky Way (MW) halo. Examples
include direct searches for weakly-interacting massive particles (WIMPs), $p$-wave suppressed
or Sommerfeld-enhanced annihilation signals, microlensing events of primordial black holes (PBHs),
{\em etc}. Most
current predictions are based on the Maxwellian approximation which is not only theoretically
inconsistent in bounded systems, but also not supported by cosmological simulations. A more
consistent method sometimes used in calculations for direct WIMP searches relies on the so-called
Eddington inversion method, which relates the DM phase-space distribution function (DF) to
its mass density profile and the total gravitational potential of the system. Originally built
upon the isotropy assumption, this method can be extended to anisotropic systems. We investigate
these inversion methods in the context of Galactic DM searches, motivated by the fact that the
MW is a strongly constrained system, and should be even more so with the ongoing Gaia survey. We
still draw conclusions that apply to the general case. In particular, we illustrate how neglecting
the radial boundary of the DM halo leads to theoretical inconsistencies. We also show that several
realistic configurations of the DM halo and the MW baryonic content entail ill-defined DFs,
significantly restricting the configuration space over which these inversion methods can apply.
We propose consistent solutions to these issues. Finally, we compute several observables
inferred from constrained Galactic mass models relevant to DM searches (WIMPs or PBHs),
{\em e.g.} moments and inverse moments of the DM speed and relative speed distributions.
}

\begin{document}
\maketitle

\section{Introduction}
\label{sec:intro}
The tremendous progress made on both direct and indirect particle dark matter (DM) searches over
the past few decades has yielded an incredible wealth of data, calling for predictions as reliable
as possible in order to draw robust conclusions on models (see reviews on DM models and search
strategies in
\eg\ Refs.~\cite{Feng2010,LavalleEtAl2012,BringmannEtAl2012c,EssigEtAl2013,Strigari2013,FreeseEtAl2013,Slatyer2016,LiuEtAl2016,CarrEtAl2016,Green2017}).
Galactic DM searches are among the most promising because the Milky Way (MW) is a
local and constrained system. However, most associated theoretical predictions are
still based on simplifying assumptions for the DM distributions in real space and/or phase space,
despite regular improvements in modeling techniques and observational constraints. Therefore,
it is usually difficult to figure out the level of uncertainties associated with these assumptions.
The fact that different studies use different assumptions makes it even more difficult to
self-consistently exploit the genuine complementarity between the constraints or discovery
avenues, which now becomes crucial as experiments have started to probe significant parts
of the parameter space allowed for popular particle DM scenarios. It is worth emphasizing that
designing constrained and theoretically sound models for the DM distribution in real space and
phase space in target systems is crucial for any astrophysical DM search, irrespective of the
DM scenario.

The Gaia mission \cite{GaiaCollaborationEtAl2016,BrownEtAl2018} is currently shedding new and
unprecedented light on the distribution of DM in the Milky Way (MW), complementary to other stellar
surveys (see \eg\ Refs.~\cite{ReidEtAl2014,PifflEtAl2014}). The Gaia data will increase the accuracy
in predictions of DM-related observables, provided reliable methods ensure their use
in a sensible way. The overall challenge is to better control not only the spatial distribution of
DM, but also its full phase-space distribution function (DF henceforth), which are the major
sources of uncertainties in predictions for DM searches. This will likely not be an easy task
\cite{Binney2017}, and there is room for significant theoretical improvement over the techniques
currently used in DM searches. The phase-space DF enters the calculations of many important
DM-related observables that depend directly on the DM velocity distribution---for example the
direct DM
detection rate, averaged $p$-wave-suppressed or Sommerfeld-enhanced annihilation cross sections,
the microlensing event rate of compact DM objects, \etc. Moreover, since the spatial distribution of
DM is the integral of the phase-space DF over momentum space, it is clear that a common framework is
necessary to make self-consistent comparisons between direct and indirect Galactic DM searches
in the broad sense, as both should exhibit some correlations (largely ignored so far, except in
a few studies, \eg\ Ref.~\cite{CerdenoEtAl2016}).

In this paper, we wish to investigate the status of some theoretical approaches that attempt to
self-consistently predict the DM phase-space DF from the full content of the target system by
virtue of the (steady-state) Boltzmann equation, the Jeans theorem, and the Poisson equation,
{\em i.e.} from first principles---we will place ourselves in the context of collisionless cold
DM from now on. These methods go beyond the simplistic approximation of a Maxwell-Boltzmann
distribution, well
suited to get fast order-of-magnitude estimates, but known for long not to apply to bounded systems
\cite{King1966}, and not to comply with dynamical constraints on the MW. These methods are
complementary to data-driven approaches (\eg\ Ref.~\cite{Herzog-ArbeitmanEtAl2018}). Other
approaches rely on fits from hydrodynamic cosmological simulations, but except for the essential
physical insight provided by generic features found in simulations
(\eg\ Refs.~\cite{PillepichEtAl2014,SchayeEtAl2015}), the
blind extrapolation of these fits to describe a single, specific, and constrained object like the
MW is questionable; not to mention the uncertainties induced by the empirical assumptions in the
description of baryonic effects and by the limited resolution. Cosmological simulations are still
very important tools to test prediction methods as they provide a framework in which all the
gravitational constituents are dynamically correlated \cite{LacroixEtAl2018}.

A well-known example and {\em a priori} self-consistent phase-space DF prediction from the
gravitational system content is the so-called Eddington inversion method \cite{Eddington1916} (and
its anisotropic extensions, like the Osipkov-Merritt models \cite{Osipkov1979,Merritt1985}), which
we discuss extensively in this paper. This approach has already been used in the context of direct
particle DM searches (see \eg\ Refs.~\cite{UllioEtAl2001,VergadosEtAl2003,
  CatenaEtAl2012,PatoEtAl2013,BhattacharjeeEtAl2013,BozorgniaEtAl2013,FornasaEtAl2014,
  LavalleEtAl2015}), as well as indirect searches (see \eg\ Refs.~\cite{FerrerEtAl2013,Hunter2014,
  BoddyEtAl2017,PetacEtAl2018}).
A net benefit from this method is that it can make use of evolved and constrained Galactic mass
models (\eg\ Refs.~\cite{CatenaEtAl2010a,McMillan2011,PifflEtAl2014,McMillan2017}), providing a
much more sensible theoretical description of the phase space. However, its validity range has not
been studied in detail in the context of DM searches, especially in a complex system like the MW,
whose gravitational potential is dominated by the baryons in the central regions. In this
work, we will show that it actually cannot apply to all DM-baryon pair configurations, leading to
ill-defined phase-space DFs even for rather conventional Galactic mass models. This is the
manifestation of gravitationally unstable DFs, and of the fact that some degrees of freedom are
missing to fully describe the system. We carefully delineate the DM-baryon parameter space where
the Eddington-like calculations may apply. We also discuss several other theoretical issues that
have been overlooked in the literature, such as the impact of the radial boundary of the system,
which should not be neglected to guarantee the existence of a closed system of equations, but may
in turn induce divergences in the velocity distribution. We propose ways to circumvent these issues,
and provide results for some observables specific to DM searches in the framework of the Galactic
mass model of Ref.~\cite{McMillan2017} (see \citeapp{app:mass_models}), namely radial profiles of
the moments of the DM speed (direct DM searches, microlensing event rate for compact DM objects,
\etc.) and of the (two-body) relative DM speed ($p$-wave-suppressed and Sommerfeld-enhanced
annihilation) distributions. We stress that although we focus on the MW in this paper, the general
aspects of this study are still relevant to the use of the Eddington formalism to describe the DM
phase-space DF of any other bounded system (with or without baryons). 
\change{We also emphasize that
  this study focuses on the theoretical self-consistency of the formalism itself, which is a first
  important step with, as we will see, quantitative consequences. It is very likely that
  several assumptions inherent to this theoretical description, like steady state, spherical
  symmetry, or the fact that potential effects coming from large substructures or recent mergers
  are neglected (\eg~the Large or Small Magellanic Cloud), will break down at some level, inducing
  another layer of systematic uncertainties. However, more detailed comparisons between the
  theoretical errors addressed here and other systematic uncertainties are left to a forthcoming
  dedicated paper\cite{LacroixEtAl2018}.\footnote{Preliminary
    results based on tests on hydrodynamic cosmological simulations show that,
    surprisingly enough, the formalism performs rather well on ``Milky Way-like''
    virtual galaxies.} }
%

The paper is organized as follows. In \citesec{sec:edd}, we review the Eddington-inversion
formalism and some of its anisotropic extensions. In \citesec{sec:issues}, we explain in detail
the issues mentioned above and their physical consequences---the divergences induced by the radial
boundary and the inability of the formalism to describe some DM-baryon configurations allowed by
kinematic constraints. In that section, we discuss some possible ways out that allow one to
recover a self-consistent description of the phase space. In \citesec{sec:dd}, we illustrate our results by calculating a series of observables relevant to particle DM direct and indirect searches. These results can be straightforwardly used for predictions in these
fields. Finally, we conclude in \citesec{sec:concl}.

\section{Eddington's inversion method and its anisotropic extensions}
\label{sec:edd}
In this section, we review the basic concepts that will be useful throughout the
discussion. Though mostly reviewing standard knowledge \cite{BinneyEtAl2008}, we will also
point out several technical details that are often overlooked or unclear in the literature.
\subsection{Jeans' theorem and spherical systems}
\label{ssec:jeans}
The Jeans theorem states that any steady-state solution of the collisionless Boltzmann equation
can be written as a function of isolating integrals of motion
\cite{Ollongren1962a,BinneyEtAl2008}. In the
particular case of a system with spherical symmetry, the energy and the modulus of the angular
momentum are such integrals of motion.
Consequently, the phase-space DF of such a system can be written
$f(\vec{r},\vec{v})\equiv f(\mathcal{E},L)$, where $L=|\vec{r}\times\vec{v}|$ is the modulus of
the angular momentum per unit mass, and
\ben
\label{eq:energy}
\mathcal{E} = \Psi(r) - \dfrac{v^{2}}{2}
\een
is the relative energy per unit mass---we assume all the DM particles in the system to be
identical. In \citeeq{eq:energy}, $v$ is the velocity, and 
\ben
\Psi(r) = \Phi_{0} - \Phi(r)
\een
is the (positive-definite) relative gravitational potential, where $\Phi(r)$ is the
solution to Poisson's equation going to 0 at infinity. The constant $\Phi_{0}$ is the value of
$\Phi(r)$ at some reference radius---usually taken to be the physical boundary of the
system---called $R_{\rm max}$ in the following.
This ensures that the potential is positive-definite over the system except at the
boundary where it vanishes. It will sometimes prove convenient to distinguish the
baryonic ($\Psi_{\rm B}$) and DM ($\Psi_{\rm D}$) contributions to the potential that we
introduce here through the following equation,
\ben
\Psi(r) =\Psi_{\rm D}(r) + \Psi_{\rm B}(r)\,.
\een
For the full system or for each component, and provided the mass profile or the density profile
are known, the relative potential $\Psi$ can be related to the mass distribution of the system
(or its individual components) through Poisson's equation, and reads
\ben
\Psi(r) = \int_{r}^{R_{\rm max}} \!  \mathrm{d}r' \, \dfrac{G m(r')}{r'^{2}} \,,
\label{eq:relative_potential}
\een
where the mass inside the sphere of radius $r$ is related to the mass density $\rho$ through
\ben
m(r) = 4 \pi \int_{0}^{r} \! \mathrm{d}r'\, \rho(r') r'^{2} \,.
\label{eq:mass}
\een
Like for the potential, the mass can be split into several components,
\eg\ a baryonic component ($m_{\rm B}$) and a DM one ($m_{\rm D}$).
We stress that the DM potential $\Psi_{\rm D}$ can be calculated from
\citeeq{eq:relative_potential} only when the DM content is specified from its density profile
$\rho$; we will see later that in some cases, we can only self-consistently get the potential
from the DF, where the radial coordinate $r$ only emerges by solving the Poisson equation
given below in \citeeq{eq:poisson}. In contrast, the baryonic potential will invariably be defined
from \citeeq{eq:relative_potential} from now on.

We limit our study to systems with spherical symmetry, therefore when dealing with a non-spherical
density component $\rho(\vec{x})$ (\eg~baryons which often have an approximate axial symmetry in
galaxies) we compute the corresponding mass inside a radius $r$ via
\ben
\label{eq:mass_axisymm}
m(r) = \int_{|\vec{x}|\le r}\mathrm{d}^{3}\vec{x} \,\rho(\vec{x})\,,
\een
and its ``spherically symmetrized potential" using \citeeq{eq:relative_potential}. This
approximation can be relaxed in principle, though the consistent treatment of an axisymmetric
distribution is much more involved (see \citesec{ssec:axisymmetry}). \change{In the following, all
  non-spherical components such as the bulge and disks in the model of Ref.~\cite{McMillan2017}
  (see App.~\ref{app:mass_models}), will be ``spherically symmetrized'' relying on
  Eq.~\eqref{eq:mass_axisymm}.}

The DF is therefore related to the mass density via
\ben
\label{eq:rho_definition}
\rho (r) = \int  \mathrm{d}^{3}\vec{v}\, f(r,\vec{v}) 
= \int \mathrm{d}^{3}\vec{v} \, f(\mathcal{E},L) \,.
\een
Note that in \citeeq{eq:rho_definition}, the DF is normalized to the total mass of the component
of interest. We keep this convention in the following.
We can further define the velocity distribution $f_{\vec{v}}$ and the speed distribution $f_{v}$
as follows:
\begin{subequations}
  \label{eq:v_df}
  \ben
  f_{\vec{v}}(\vec{v},r) &\equiv& \frac{f(\mathcal{E},L)}{\rho (r)}\\
  f_{v}(|\vec{v}|,r) &\equiv& v^2 \int {\rm d}\Omega_v \, f_{\vec{v}}(\vec{v},r)\,,
  \een
\end{subequations}
where ${\rm d}\Omega_v$ encodes the angular content of the velocity distribution. From the
above definition, both $f_{\vec{v}}$ and $f_{v}$ carry the usual units and are normalized
to unity. We stress that the DF introduced in \citeeq{eq:rho_definition} is implicitly assumed to
further satisfy Poisson's equation
\ben
\Delta \Psi_i = -4\,\pi \, G \,\rho_i(r) = -4\,\pi\, G \int \mathrm{d}^{3}\vec{v} \,
f_i({\cal E},L)
= -4\,\pi\, G \int_0^\Psi {\rm d}{\cal E}'\,\sqrt{2(\Psi-{\cal E}')}\,f_i({\cal E}',L)
\,,
\label{eq:poisson}
\een
which will turn out to be important later on. When $\rho_i(r)$ is specified, the above
equation reduces to \citeeq{eq:relative_potential} if the boundary condition $\Psi_i(R_{\rm max})=0$
is considered (here, this will always be the case for the baryonic component). Otherwise,
\citeeq{eq:poisson} will have to be solved explicitly to compute the potential.
The $i$ index makes
it clear that although the energy ${\cal E}$ depends on the full potential $\Psi=
\sum_i\Psi_i$ and thereby on all the gravitational components of the system, the Poisson
equation only relates the individual components to their own phase-space DF.

There is no general classification of the solutions of the collisionless Boltzmann equation.
Therefore, further assumptions on the properties of the phase space are needed. In the following,
we recall the main equations of the Eddington inversion formalism---which allows one to derive a
phase-space DF for a given galactic mass model and for particular assumptions on the anisotropy of
the system---before discussing in detail theoretical issues that may arise from the
method.

\subsection{Eddington's inversion for an isotropic system}
\label{ssec:iso}
We first set about describing the simplest case of a spherically symmetric and isotropic
DM distribution. In that case, the angular momentum is irrelevant, and the dependence of the DF
on integrals of motion simplifies to an energy dependence, $f \equiv f(\mathcal{E})$. Such a DF
is referred to as ergodic. Using \citeeq{eq:energy} as a change of variables to eliminate the
velocity, we can rewrite \citeeq{eq:rho_definition} as
\ben
\label{eq:rho_fE}
\rho (r) = 4 \pi \sqrt{2} \int_{0}^{\Psi(r)} \!
f(\mathcal{E}) \sqrt{\Psi(r) - \mathcal{E}} \, \mathrm{d}\mathcal{E}\,.
\een
Note that we only consider \emph{self-gravitating} systems, which means all particles in the
system are gravitationally bound to it and have $\mathcal{E} \geq 0$. As a result,
$f(\mathcal{E} < 0) = 0$. This translates into a lower bound of $\mathcal{E}=0$ in the integral
in \citeeq{eq:rho_fE}. For general systems that are not self-gravitating, the lower bound would be
$\mathcal{E}=-\infty$.

Since $\Psi$ is a monotonically decreasing function of $r$ in a realistic stationary system, one can
define $\rho$ as a function of $\Psi$ instead of $r$. Differentiating \citeeq{eq:rho_fE} with
respect to $\Psi$, one obtains
\ben
\dfrac{\mathrm{d}\rho}{\mathrm{d}\Psi} = \sqrt{8}\pi \int_{0}^{\Psi} \! \dfrac{f(\mathcal{E})}{\sqrt{\Psi - \mathcal{E}}} \, \mathrm{d}\mathcal{E}\,.
\label{eq:abel_equation_sec2}
\een
This is an Abel equation, which can be inverted to give Eddington's formula
\cite{Eddington1916,BinneyEtAl2008}:
\ben
f(\mathcal{E}) = \dfrac{1}{\sqrt{8}\pi^{2}} \dfrac{\mathrm{d}}{\mathrm{d}\mathcal{E}}
\int_{0}^{\mathcal{E}} \! \dfrac{\mathrm{d}\Psi}{\sqrt{\mathcal{E} - \Psi}}\,
\dfrac{\mathrm{d}\rho}{\mathrm{d}\Psi} \,.
\label{eq:eddington_form1}
\een
A more convenient form of Eddington's formula that does not explicitly feature a derivative with
respect to $\mathcal{E}$ can be obtained after integrating by parts:
\ben
\label{eq:eddington_form2}
f(\mathcal{E}) &=& \dfrac{1}{\sqrt{8}\pi^{2}}
\left\{ \dfrac{1}{\sqrt{\mathcal{E}}} \left[ \dfrac{\mathrm{d}\rho}{\mathrm{d}\Psi} \right]_{\Psi=0}
+ \int_{0}^{\mathcal{E}} \! \dfrac{\mathrm{d}\Psi}{\sqrt{\mathcal{E} - \Psi}} \,
\dfrac{\mathrm{d}^{2}\rho}{\mathrm{d}\Psi^{2}}  \right\}\\
\Big[  &=& \dfrac{2}{\sqrt{8}\pi^{2}}
  \left\{ \dfrac{1}{2\,\sqrt{\mathcal{E}}}
  \left[ \dfrac{\mathrm{d}\rho}{\mathrm{d}\Psi} \right]_{\Psi=0}
  + \sqrt{\cal E} \left[ \dfrac{{\rm d}^2\rho}{{\rm d}\Psi^2} \right]_{\Psi=0}
  + \int_{0}^{\mathcal{E}} \! {\rm d}\Psi \,\sqrt{\mathcal{E} - \Psi}
  \, \dfrac{\mathrm{d}^{3}\rho}{\mathrm{d}\Psi^{3}} \,
     \right\}  \Big]\,.\nn
\een
This is the form we will use and discuss extensively in the following (the last line in brackets
corresponds to an additional integration by parts, which will prove insightful later on).
Integrating \citeeq{eq:abel_equation_sec2}, one can reconstruct the density profile from the DF:
\ben
\rho (\Psi) = \rho(\Psi = 0) + 4 \pi \sqrt{2} \int_{0}^{\Psi} \! \mathrm{d}\mathcal{E}\,
\sqrt{\Psi - \mathcal{E}} \, f(\mathcal{E})\,,
\label{eq:rho_reconstruction}
\een
where $\rho(\Psi = 0) = \rho(r=R_{\mathrm{max}})$ is the density at the boundary of the system,
very often neglected in the literature whereas it is an important ingredient to test the
self-consistency of the chain of calculations (one must obviously recover the initial input
density profile from integrating the DF). Indeed, the Abel inversion is performed on
$\mathrm{d}\rho/\mathrm{d}\Psi$ rather than $\rho$. We also emphasize the importance
of the term $\propto 1/\sqrt{\cal E}$ in \citeeq{eq:eddington_form2} to get a consistent
reconstruction of $\rho$ up to the radial boundary $R_{\rm max}$ of the system, except
in the special limit $R_{\rm max}\to \infty$. As a potentially important technical consequence,
the self-consistent normalization of the velocity or speed distributions $f_{\vec{v}/v}$ defined
in \citeeq{eq:v_df} is no longer guaranteed---neglecting the term $\propto 1/\sqrt{\cal E}$
therefore imposes to normalize the distributions $f_{\vec{v}/v}$ by hand. An illustration is
presented in \citefig{fig:radial_cut}, where the dashed curves show the reconstructed profiles
when neglecting $\rho(\Psi = 0)$, the dotted curves further neglect the term of the DF
$\propto 1/\sqrt{\cal E}$, all compared with the fully reconstructed profiles (solid lines).

\change{As a side remark, note that $\rho$ and $\Psi$ need not be related for the Eddington
  inversion to work. For instance, if the system contains DM and baryons, $\rho$ refers to the DM
  density, whereas $\Psi = \Psi_{\mathrm{D}} + \Psi_{\mathrm{B}}$ is the total potential. In that case,
  $\Psi$ cannot be determined from the sole knowledge of the DM density. That $\rho$ and $\Psi$
  can be independent will have consequences in terms of physical self-consistency of the derived
  DF, as will be discussed in \citesec{ssec:gamma}.}

\begin{figure*}[!t]
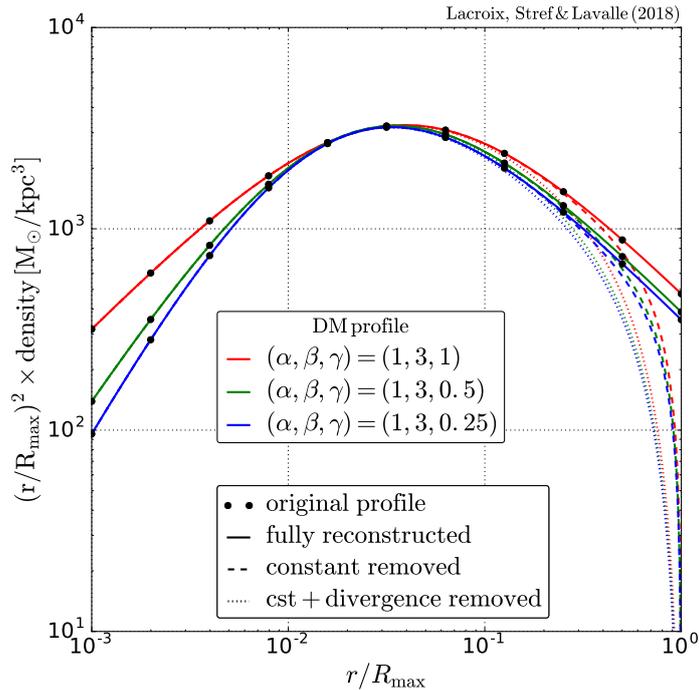

\centering
\includegraphics[width=0.6\linewidth]{{{fig_density_mcmillan_2}}}
\caption{\small
  Initial density profiles with different
  inner slopes $\gamma=0.25,0.5,1$ (see \citeapp{app:mass_models}) taken from
  Ref.~\cite{McMillan2017} and their reconstruction from \citeeq{eq:rho_reconstruction}. The
  black circles show the original profiles, the solid lines the full reconstructions based
  on \citeeq{eq:rho_reconstruction}, the dashed lines neglect the constant term of
  \citeeq{eq:rho_reconstruction}, and the dotted lines neglect both the latter and the
  $1/\sqrt{\cal E}$ term in the calculation of the DF
  $f({\cal E})$ [see \citeeq{eq:eddington_form2} and \citesec{ssec:rmax}].
}
\label{fig:radial_cut}
\end{figure*}

\subsection{Anisotropic extensions}
\label{ssec:aniso}
When the system features some degree of anisotropy, the density profile and the total
gravitational potential are no longer sufficient to determine the DF because the angular momentum
$\vec{L}$ enters the game, and an ansatz for $f(\mathcal{E},\vec{L})$ is required to account for
the dependence of the DF on these new degrees of freedom---for the spherically symmetric systems
considered here, the phase space is only extended by the modulus $|\vec{L}|=L$. An anisotropic
system is usually characterized in terms of an anisotropy parameter \cite{Binney1980}:
\ben
\beta(r) = 1 - \dfrac{\sigma_{\theta}^{2} + \sigma_{\phi}^{2}}{2 \sigma_{r}^{2}}\,,
\een
where $\sigma_{r}$, $\sigma_{\theta}$ and $\sigma_{\phi}$ are the velocity dispersions in spherical
coordinates. If orbits in the system of interest are mostly tangential, we have
$\sigma^{2}_{r}\ll\sigma^{2}_{\theta}+\sigma^{2}_{\phi}$ and $\beta<0,\,|\beta|\gg1$. If orbits are
mostly radial, we get $\sigma^{2}_{r}\gg\sigma^{2}_{\theta}+\sigma^{2}_{\phi}$ and $\beta=1$.
In the following, we describe two simple ans\"atze that provide semi-analytical solutions
from the Abel inversion procedure in the anisotropic case, and briefly discuss more sophisticated
approaches.

\subsubsection{Constant anisotropy}
A simple extension of the Eddington method deals with systems having a constant anisotropy
parameter $\beta(r)=\beta_{0}$.
The simplest ansatz for the DF that separates the effects of energy and angular momentum
takes the following form \cite{Henon1973,KentEtAl1982,BinneyEtAl2008}:
\ben
f_{\beta_{0}}(\mathcal{E},L) = G(\mathcal{E}) L^{-2\beta_{0}}\,.
\label{eq:df_constant_beta}
\een
The function $G$ is related to the density profile through
\ben
\chi \equiv r^{2\beta_{0}}\rho &= \lambda(\beta_{0})
&
\int_{0}^{\Psi} \! G(\mathcal{E})\,(\Psi-\mathcal{E})^{\frac{1}{2}-\beta_{0}} \,
\mathrm{d}\mathcal{E}\,,
\label{eq:density_beta}
\een
where
\ben
\lambda(\beta_{0}) = 2^{\frac{3}{2}-\beta_{0}}\pi^{\frac{3}{2}}
\frac{\Gamma(1-\beta_{0})}{\Gamma(3/2-\beta_{0})}\,,
\een
where $\Gamma$ is the Gamma function (Euler integral of the second kind).
This leads to the Abel equation
\ben
\frac{\mathrm{d}^{n}\chi}{\mathrm{d}\Psi^{n}} = \lambda(\beta_{0})\left(\frac{1}{2}-\beta_{0}\right)!
\int_{0}^{\Psi} \! G(\mathcal{E})(\Psi-\mathcal{E})^{\frac{1}{2}-\beta_{0}-n} \, \mathrm{d}\mathcal{E},
\label{eq:abel_eq_constant_beta}
\een
where
\ben
& \left(\dfrac{1}{2}-\beta_{0}\right)! 
& \equiv \left\{ 
\begin{array}{ll}
\left( \dfrac{1}{2}-\beta_{0} \right)  ... \left( \dfrac{1}{2}-\beta_{0}-(n-1) \right) \ &\mathrm{for}\ \beta_{0}<\dfrac{1}{2} \\
1\ &\mathrm{for}\ \dfrac{1}{2}\leqslant \beta_{0}<1 
\end{array}
\right.,
\een
and
\ben
n=\left[\frac{3}{2}-\beta_{0}\right]\,,
\label{eq:index_beta}
\een
with $[x]$ the floor of $x$. The solution of this equation can be expressed as
\ben
G(\mathcal{E}) = 
\dfrac{\sin((n-1/2+\beta_{0})\pi)}{\pi \lambda(\beta_{0})\left(1/2-\beta_{0}\right)!} 
\frac{\mathrm{d}}{\mathrm{d}\mathcal{E}}\int_0^{\mathcal{E}} \! \mathrm{d}\Psi\,
\frac{\mathrm{d}^{n}\chi}{\mathrm{d}\Psi^{n}}(\mathcal{E}-\Psi)^{n-3/2+\beta_{0}}\,,
\label{eq:G_constant_beta}
\een
We note that in the isotropic limit $\beta_{0}\rightarrow0$, the expression of $G$ in
\citeeq{eq:G_constant_beta} boils down to the Eddington DF given in \citeeq{eq:eddington_form1}
as expected.
\change{If $\beta_{0}$ is a half-integer, the integral in Eq.~(\ref{eq:G_constant_beta}) boils down to a derivative \citep{Cuddeford1991}. This allows one to analytically express the DF of any system with a half-integer anisotropy \citep{Evans2005}.}

\subsubsection{Osipkov-Merritt model}
Another extension of the Eddington formalism is the Osipkov-Merritt DF
\cite{Osipkov1979,Merritt1985} which describes a system where the anisotropy parameter is
no longer constant, but takes the following radial dependence:
\ben
\beta(r) = \frac{r^2}{r^2+r_{\rm a}^2}\,,
\label{eq:beta_om}
\een
where $r_{\rm a}$ is a free parameter referred to as the anisotropy radius.
This model is isotropic in the inner regions $r\ll r_{\rm a}$, while it exhibits a full
radial anisotropy in regions $r\gg r_{\rm a}$. It cannot describe tangential anisotropy. The full
isotropic case is recovered in the limit $r_{\rm a}\to\infty$. This expression is derived by
assuming that the DF no longer factorizes out its dependence on energy and angular momentum, but
mixes them through a variable $Q$, 
\ben
f(\mathcal{E},L) = f_{\mathrm{OM}}(Q)\,,
\label{eq:df_om}
\een
where $Q = \mathcal{E} - \dfrac{L^{2}}{2 r_{\mathrm{a}}^{2}}$. By solving
\ben
\rho(r) = \int{\rm d}^3\vec{v} \, f_{\mathrm{OM}}(Q)\,,
\een
one readily obtains
\ben
\rho(r) = \frac{r_{\rm a}^2}{r^2+r_{\rm a}^2}\,\rho_{\rm OM}(r)\,,
\label{eq:density_om}
\een
where
\ben
\rho_{\rm OM}(r) =  \rho_{\rm OM}\left(\Psi(r) \right) =
4 \pi \sqrt{2} \int_{0}^{\Psi}f_{\rm OM}(Q) \sqrt{\Psi - Q} \, \mathrm{d}Q\,.
\een
The Abel equation is then
\ben
\frac{\mathrm{d}\rho_{\rm OM}}{\mathrm{d}\Psi} =
\sqrt{8}\pi\int_{0}^{\Psi}\frac{f_{\rm OM}}{\sqrt{\Psi-Q}}\,\mathrm{d}Q\,,
\een
and its solution
\ben
f_{\rm OM}(Q) = \dfrac{1}{\sqrt{8}\pi^{2}}
\frac{\mathrm{d}}{\mathrm{d}Q}\int_{0}^{Q}\frac{\mathrm{d}\Psi}{\sqrt{Q-\Psi}}
\frac{\mathrm{d}\rho_{\rm OM}}{\mathrm{d}\Psi}\,.
\label{eq:om_form1}
\een
The expression of $f_{\rm OM}$ is identical to that of the standard Eddington DF in
\citeeq{eq:eddington_form2} when $Q$ and $\rho_{\rm OM}$ are identified with
$\mathcal{E}$ and $\rho$, respectively (in the isotropic limit $r_{\rm a}\rightarrow\infty$, both
expressions match).
\subsubsection{Other possibilities}
\label{sssec:}
The two methods discussed above are the simplest ones accounting for anisotropy in the
velocity distribution, as they depend only on one free parameter ($\beta_{0}$ or $r_{\rm a}$). Other
DFs involving more free parameters can be found in the literature, such as a straightforward
generalization of both constant anisotropy and Osipkov-Merritt models \cite{Cuddeford1991},
\ben
f(\mathcal{E},L) = G(Q)L^{-2\beta_{0}}\,.
\een
Motivated by the anisotropy profiles $\beta(r)$ observed in N-body simulations, some authors have
also considered linear combinations of the constant anisotropy DF and the Osipkov-Merritt
DF \cite{BozorgniaEtAl2013}
\ben
f(\mathcal{E},L) = w f_{\rm OM}(Q)+(1-w)G(\mathcal{E})L^{-2\beta_{0}}\,,
\een
while others have looked at different functional forms \cite{WojtakEtAl2008}:
\ben
f(\mathcal{E},L) = F(\mathcal{E})
\left(1+\frac{L^{2}}{2L_{0}^{2}}\right)^{-\beta_{\infty}+\beta_{0}}L^{-2\beta_{0}}\,.
\een
Models of Refs.~\cite{BozorgniaEtAl2013,WojtakEtAl2008} both contain a set of three free parameters
($\{w,r_{\rm a},\beta_{0}\}$ or $\{L_{0},\beta_{0},\beta_{\infty}\}$) calibrated on simulations.
Irrespective of the different proposals to cope with anisotropy in the DM velocity field,
we stress that the latter is still hardly constrained by kinematic observations of visible
matter.

\subsection{Beyond spherical symmetry}
\label{ssec:axisymmetry}

\change{In this study, we will not go beyond spherical symmetry except to approximately integrate
  the effects of some non-spherical components like the baryonic bulge and disks [see
    \citeeq{eq:mass_axisymm} and discussion below]. Here, for the sake of completeness, we just
  review some more involved theoretical methods that can be used to cope with this delicate
  problem}.
When dealing with a system that is not spherically symmetric, the energy and the angular momentum
might not be the most convenient variables to work with. The authors of
Refs.~\cite{BinneyEtAl2008,SandersEtAl2016} promote instead the \textit{angle-action}
variables as a phase-space coordinate system. The components of the action vector $\vec{J}$ are
integrals of motion and the angle vector $\vec{\Theta}$ is the Hamiltonian conjugate of $\vec{J}$.
A crucial property of the actions is their constancy in a slowly varying potential. In such a
potential, a DF of the form $f(\vec{J})$ is then also a constant. This property was used
as a starting point in Ref.~\cite{PifflEtAl2015} to compute a phase-space model
of the Milky Way, assuming baryons are slowly accreted onto an initially spherical dark halo. This
led, in this theoretical framework, to the exclusion of an adiabatic compression of the
dark halo \cite{BinneyEtAl2015}, favoring instead heating at its center and the presence of a
$\sim$2 kpc core \cite{ColeEtAl2017} in agreement with a detailed study of the bar/bulge dynamics
\cite{PortailEtAl2017}. 

The philosophy behind this technique is opposite to Eddington's since here the starting point is
the DF, from which the potential is computed through an iterative procedure, while in the Eddington
case one starts with the potential and the density and derives the DF from there. Just like there
is a lot of freedom when choosing the functional form of the DF in the anisotropic extensions of the
Eddington inversion method, there is also some freedom in choosing the functional form of the
action-dependent DF. Assumptions must therefore be made on its dependency with respect to each
action and this may introduce theoretical uncertainties in the calculation which are difficult to
evaluate. Nevertheless, this method constitutes the state of the art of Galactic phase-space
modeling and it captures details beyond the reach of the Eddington formalism.

This level of detail might not be required in the context of DM searches though, as one is mostly
interested in evaluating the astrophysical uncertainties relevant to complementary
observables of interest in a self-consistent framework. The Eddington formalism actually provides
such a framework, while being in practice more flexible than the angle-action approach. Moreover,
global dynamical constraints are easier to account for in the Eddington approach from a technical
point of view. However, as we will show in the following, the Eddington inversion is not a
self-regulated approach as it does not prevent from getting unstable or ill-defined phase-space
configurations, whereas action-angle methods are a priori immune to these defects. It is therefore
important to delineate as rigorously as possible the domain of application of the Eddington
inversion. Ultimately, more systematic comparisons with action-angle methods should help further
reduce the theoretical uncertainties and provide complementary understanding of the potential
failures of the Eddington inversion, but this goes beyond the scope of this paper.

\section{Some issues of the Eddington formalism}
\label{sec:issues}
In this section, we discuss in detail two issues that we have identified in the Eddington
inversion method, and which have been overlooked in the literature focused on DM searches.
The first one concerns the impact of the spatial boundary of the dark halo, which is usually
neglected while this leads to theoretical inconsistency and also potentially to mistreatments
of the tail of the DM velocity distribution. The second one is related to the fact that some 
perfectly licit DM-baryons configurations may actually lead to unstable DFs.

We recall that the main benefits of the Eddington formalism (including its anisotropic
extensions) in the context of DM searches is precisely to provide a self-consistent and
constrained framework to compute both density-dependent and velocity-dependent observables.
A noticeable strength is to be able to use a kinematically constrained Galactic mass model and
self-consistently propagate the associated uncertainties to the DM observables. However, the two
issues mentioned above and further detailed in this section jeopardize this possibility.
  
In the following, all concrete calculations of the DFs will be made using the best-fit
Galactic mass models of Ref.~\cite{McMillan2017} (McM17 models henceforth), unless specified
otherwise. The nominal model is featured by an NFW DM halo and a baryonic component
made of a stellar bulge, stellar disks, and gaseous disks, all of these components being
constrained from recent kinematic data. The parameters of these models are summarized in
\citeapp{app:mass_models}.
\subsection{Radial-boundary-induced divergence, the escape speed, and some regularization
  procedures}
\label{ssec:rmax}
\subsubsection{Characterization of the spatial-boundary-induced divergence}
\label{sssec:div_rmax}
A generic issue with the Eddington DF is the presence of a divergence in the limit
$\mathcal{E}\to 0$ due to the term
$(\mathrm{d}\rho/\mathrm{d}\Psi)_{\Psi=0}\times 1/\sqrt{\mathcal{E}}$ present in
\citeeq{eq:eddington_form2}. This derivative is evaluated at the radial boundary of the
system and does not vanish for conventional halo profiles, unless the boundary is sent
to infinity. However, this boundary must be finite just because of the presence of neighboring
galaxies. It is precisely what allows us to make a realistic interpretation of an escape speed,
which has some impact on \eg\ direct searches of low-mass WIMPs \cite{LavalleEtAl2015}.

\begin{figure*}[!t]
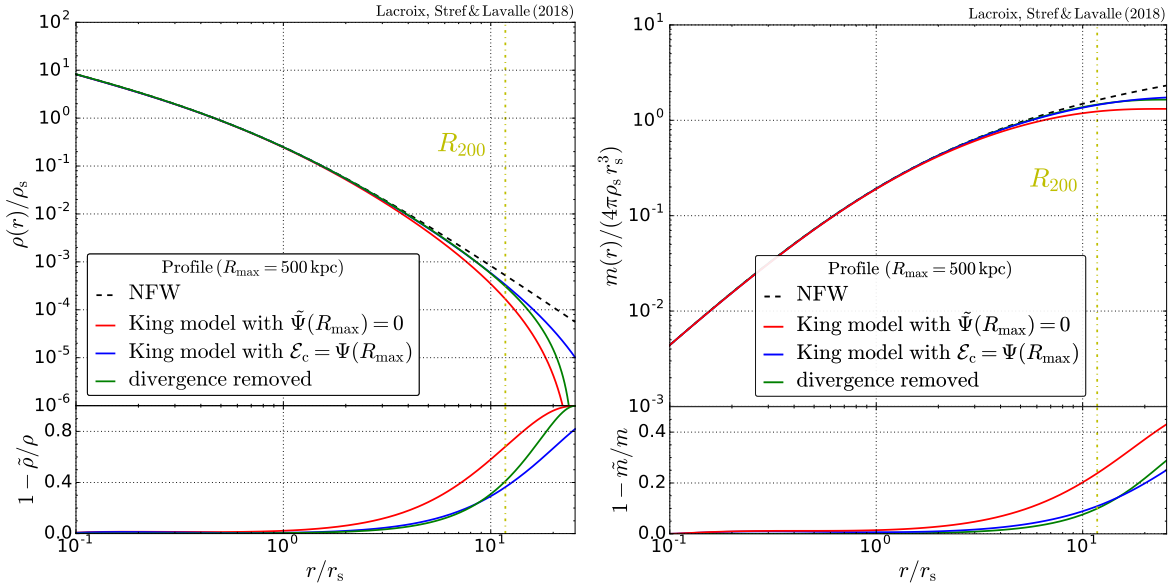

\centering
\includegraphics[width=0.495\linewidth]{{{fig_reconstructed_density_king_models}}}
\includegraphics[width=0.495\linewidth]{{{fig_reconstructed_mass_king}}}
\caption{\small Same as \citefig{fig:reconstructed_rho}, showing the profiles resulting from DFs
  regularized \`a la King, based on the ansatz of \citeeq{eq:king_model}. {\bf Left panel:}
  reconstructed density profiles compared with the initial one. {\bf Right panel:} corresponding
  dark halo mass profiles.}
\label{fig:reconstructed_rho_king}
\end{figure*}

This diverging term $\propto 1/\sqrt{\mathcal{E}}$ is actually very often dropped without deep
justifications. \change{However, this jeopardizes the self-consistency of the approach, since the
reconstructed DM density profile then significantly departs from the initial one, unless one is
interested in describing only the inner parts of the Galaxy (see \citefig{fig:radial_cut} and the
green curve in Fig.~\ref{fig:reconstructed_rho_king}, as well as a more extended discussion on the
density profile in \citesec{sssec:reg_dens}). More specifically, the reconstructed density differs
from the initial one by $\sim 10\%$ above $0.1 R_{\mathrm{max}}$, i.e.~$\sim 2 r_{\mathrm{s}}$, and the
difference increases even more at larger radii. Even if these numbers do not look dramatic,
they still undermine the appealing aspects of this framework as a consistent and global framework
for DM-signal predictions, as one loses control on the input mass model uncertainties}. On the
other hand, sending the boundary to infinity spoils control on the tail of the velocity
distribution.

Since the speed distribution $f_{v}(v,r)$ is directly related to the DF through \citeeq{eq:v_df},
the divergence when $\mathcal{E}\to 0$ translates into a divergence in velocity space when
$v^2\to 2\psi(r)$, \ie\ at the escape speed and at \textit{any} position in the system.
In \citefig{fig:divergence_vesc} (left panel), we illustrate this divergence in the speed
distribution evaluated at a radius $r=20\,\rm kpc$ (solid red line) in our default halo model
with a radial extension set to $R_{\rm max}=500\,\rm kpc$. This divergence is the sign that the
system under consideration is artificially compressed in phase space. A population of
particles is squeezed near the escape speed, while we would expect a stable DF
to verify $f(\mathcal{E}\to0)\to 0$. The right panel of \citefig{fig:divergence_vesc} shows
the pathological DF $f(\mathcal{E})$ as a function of $\mathcal{E}$ (solid red curve), where
the divergence occurs at $\mathcal{E}\to 0$.

This divergence is present whenever the derivative $(\mathrm{d}\rho/\mathrm{d}\Psi)_{\Psi=0}$ is
non-zero, which is always the case for conventional halo profiles with finite boundaries unless
one modifies the asymptotic behavior at the boundaries. This issue is therefore intimately
related to the spatial extension of the system, since the troublesome derivative is evaluated
at $\Psi=0$ (equivalently $r=R_{\rm max})$. The gravitational potential being defined up to a
constant, the position $r=R_{\rm max}$ where $\Psi$ vanishes is a matter of choice. For example,
taking $R_{\rm max}\to\infty$ solves the issue and the DF satisfies $f(\mathcal{E}\to 0)\to 0$,
as shown by the blue solid curve in the right panel of \citefig{fig:divergence_vesc}.
This actually matches with the boundary condition of having the gravitational potential
$\psi(r)=-\phi(r) \to 0$ as $r\to\infty$ when solving the Poisson equation. The speed
distribution is then regularized---see the blue solid curve in the left panel of
\citefig{fig:divergence_vesc}. The DF obtained for this idealized--though
unrealistic--choice of $R_{\mathrm{max}}$ is fully consistent with the mass model and is a solution
of the collisionless Boltzmann equation by construction. 

This leads to the following interpretation of the divergence showing up at finite radial
extensions: particles that could have probed infinite distances in agreement with the
conventional infinite boundary condition are now prevented from radially escaping the system and
have their phase space compressed accordingly.

However, choosing $R_{\rm max}\to\infty$ is physically problematic in this context. DM halos
always have a finite extension due to the gravitational influence of other neighboring halos
(like the dark halo of M31 in the case of our Galaxy), or the host halo if the system under
consideration is a subhalo. Taking this finite extension into account is crucial for DM searches
as its fixes the definition of the escape speed of the system
\ben
v_{\rm esc}(r) = \sqrt{2(\phi(R_{\rm max})-\phi(r))}\,.
\een
The value of the escape speed at the position of the Solar System is for instance a major
ingredient when making predictions for direct WIMP searches in the low-mass region
\cite{LavalleEtAl2015}. The escape speed is also a target observable that can be inferred
from stellar surveys \cite{PifflEtAl2014a,Herzog-ArbeitmanEtAl2018}. Finally, in the particular
case of the MW, the closest neighbor is the Andromeda galaxy which is about $800\,\rm kpc$ away
from the Galactic center. Consequently the Galactic halo cannot extend much farther than
$R_{\rm max}\sim 500\,\rm kpc$, which we take as our reference value from now on.
\footnote{Note that $R_{\rm max}$ is almost twice as large as the estimated virial radius
  $R_{200}\sim 250\,\rm kpc$.}

In the left panel of \citefig{fig:divergence_vesc_OM}, we compute the relative change in the
escape speed when increasing the value of $R_{\rm max}$. One can see that the escape speed at
$r=8\,\rm kpc$ increases by up to 10\% ($\sim 50$km/s) when the radial boundary moves further out.
The relative increase gets bigger as the position is farther away from the center of the halo,
though lower in absolute value. It is therefore important to be as consistent as possible
when one wants to relate the concept of escape to the phase-space DF.

The discussion above focused on the isotropic case, but the situation is very similar in the
anisotropic case with a constant anisotropy parameter $\beta$. Sending $R_{\rm max}$ to infinity
removes the diverging term in $G(\mathcal{E})$ [see \citeeq{eq:df_constant_beta}], and the DF is
regularized at the cost of changing the escape speed. However, the situation is different in the
Osipkov-Merritt case, as the troublesome derivative is
$(\mathrm{d}\rho_{\rm OM}/\mathrm{d}\Psi)_{\Psi=0}$ with $\rho_{\rm OM}$ defined in
\citeeq{eq:density_om}. One can check that if the density behaves as a power-law at
large radii $\rho\propto r^{-b}$---as is almost always the case---then
$\mathrm{d}\rho_{\rm OM}/\mathrm{d}r\propto r^{1-b}$ and $\mathrm{d}\Psi/\mathrm{d}r\propto r^{1-b}$,
meaning that the derivative $(\mathrm{d}\rho_{\rm OM}/\mathrm{d}\Psi)_{\Psi=0}$ goes to a constant
as $R_{\rm max}$ goes to infinity. Consequently, in the Osipkov-Merritt model, not only
does considering an infinite system affect the escape speed, but it also does not remove the
phase-space divergence, as illustrated in the right panel of \citefig{fig:divergence_vesc_OM}.
Moreover, for
this model the divergence in the speed distribution (Eq.~\ref{eq:v_df}) does not
occur at $v_{\mathrm{esc}}$ but appears in the peak of the distribution due to the angular integral.
This makes it more difficult to regularize the DF.

In the following, we discuss different ways of getting rid of this divergence in order to
obtain physically viable solutions.

\begin{figure*}[!t]
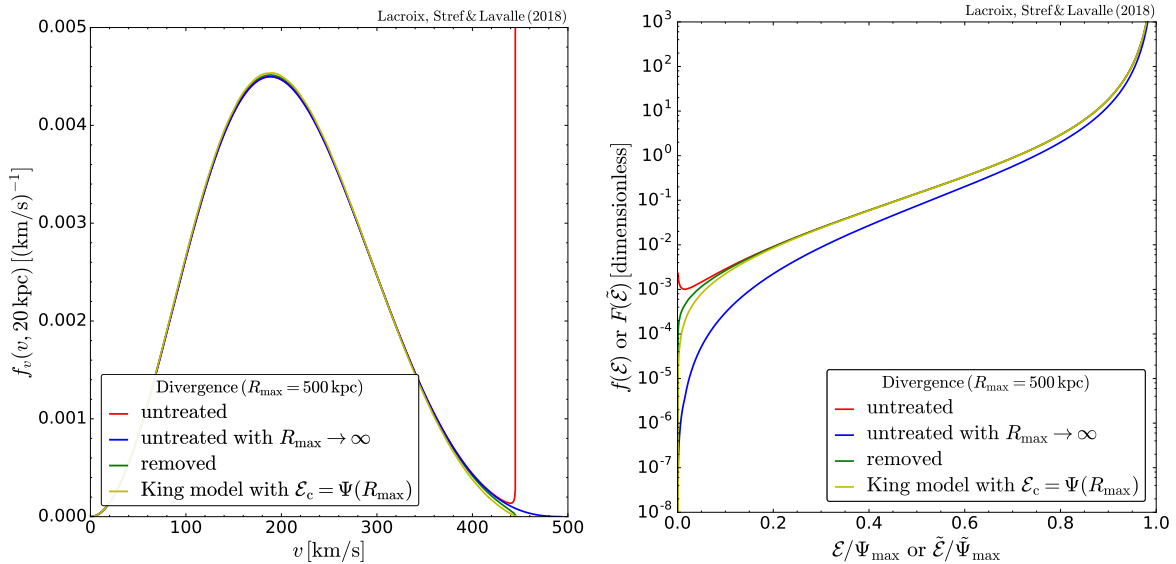

\centering
\includegraphics[width=0.500\linewidth]{{{fig_speed_distribution_20kpc_divergence}}}
\includegraphics[width=0.49\linewidth]{{{fig_f_eps_tail_v2}}}
\caption{\small \textbf{Left panel:} Velocity distribution functions (i) for the NFW profile of
  Ref.~\cite{McMillan2017} at a radius of 20 kpc, for three situations regarding the status of the
  divergence at the escape velocity; and (ii) for the regularization \`a la King given
  in \citeeq{eq:king_model}. The red, blue, green, and yellow lines represent the DFs obtained
  by keeping the divergence, sending the boundary of the system to infinity, removing the
  divergence, and using the regularization \`a la King, respectively.
  \textbf{Right panel:} Corresponding DFs as a function of the energy
  ${\cal E}$. DFs are in units of $\rho_{\rm s}(4\,\pi\,G_{\rm N}\,\rho_{\rm s}\,r_{\rm s}^2)^{-3/2}$.}
\label{fig:divergence_vesc}
\end{figure*}

\begin{figure}[!t]
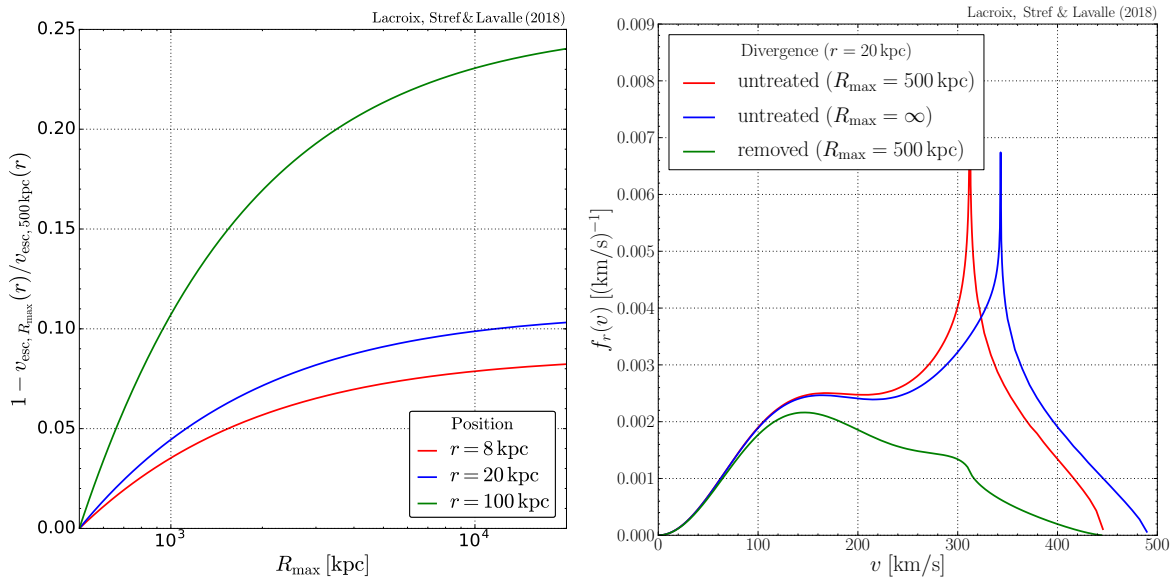

\centering
\includegraphics[width=0.485\linewidth]{{{fig_err_vesc_rmax}}}
\includegraphics[width=0.505\linewidth]{{{fig_fv_OM_20kpc_divergence_treatment}}}
\caption{\small \textbf{Left panel:} Relative variation of the escape velocity at a position $r$
  as the system's boundary $R_{\rm max}$ is modified (the reference value is set to
  $R_{\rm max}=500\,\rm kpc$). \textbf{Right panel:} Same as left panel of
  \citefig{fig:divergence_vesc}, for the Osipkov-Merritt model.}
\label{fig:divergence_vesc_OM}
\end{figure}

\subsubsection{Regularization through the density profile}
\label{sssec:reg_dens}
The most simple solution to the boundary-induced divergence is to slightly modify the
DM density profile in such a way that it is still consistent with the kinematic
constraints which it was derived from, and that $(\mathrm{d}\rho/\mathrm{d}r)_{r=R_{\rm max}}$
vanishes---this is not the case with standard NFW, $\alpha\beta\gamma$, or Einasto profiles. If
such a solution exists, then the Eddington formalism fully applies and provides a self-consistent
description of the phase space up to the spatial boundary of the DM halo.
Therefore, since the bulk of the kinematic constraints pertains to the inner 50 kpc of the
Galaxy \cite{McMillan2017}, we need to make sure that both the input DM mass profile and
gravitational potential are not affected in this range. The kinematics of satellite galaxies can
also be used to constrain the MW mass within $\sim 300$ kpc with larger uncertainties
(see \eg\ Ref.~\cite{WatkinsEtAl2010}), but not farther, so the modified mass profile should not
depart too much from the initial one and remain consistent with these bounds.

Modifications of standard functional forms of the density profile can be found in the literature.
For instance, the authors of Ref.~\cite{SpringelEtAl1999} (see also
Ref.~\cite{KazantzidisEtAl2004}) account for tidal stripping of the outer parts of a halo with an
exponential suppression. However, this modified profile has
$(\mathrm{d}\rho/\mathrm{d}\Psi)_{\Psi=0}\neq0$ and therefore leads to a diverging DF once we set
a radial boundary. Instead, we propose the following alternative density profile to model the
halo:  
\ben
\tilde{\rho} = \rho-\Psi_{\rm D}\left(\frac{\mathrm{d}\rho}{\mathrm{d}\Psi_{\rm D}}\right)_{\Psi=0}.
\label{eq:alt_density}
\een
The corresponding DM component, the gravitational potential of which is $\Psi_{\rm D}$, is
consistently obtained from the Poisson equation (with the vanishing condition at the radial
boundary $R_{\rm max}$), which then reduces to
\ben
\label{eq:alt_potential}
\tilde{\Psi}_{\rm D}(r) &=& G\int_r^{R_{\rm max}}dr'\,\frac{\tilde{m}_{\rm D}(r')}{r'^2}\\
{\rm where}\;\tilde{m}_{\rm D}(r) &=&4\,\pi\int_0^r dr'\,r'^2\,\tilde{\rho}(r')\,.\nn
\een
This potential is the one to be used in the Eddington inversion along with the modified
density profile $\tilde{\rho}$ (the baryonic component is left unchanged). That new
density profile $\tilde{\rho}$, defined
in \citeeq{eq:alt_density}, flattens at the edge of the system, \ie\
\ben
(\mathrm{d}\tilde{\rho}/\mathrm{d}\tilde{\Psi})_{\tilde{\Psi}=0} =
    [(\mathrm{d}\tilde{\rho}/\mathrm{d}r)
      (\mathrm{d}\tilde{\Psi}/\mathrm{d}r)^{-1}]_{r = R_{\mathrm{max}}} = 0\,.\nn
\een
This flattening at $r\to R_{\rm max}$ can be thought of as the border with the
homogeneous background or with neighboring self-gravitating systems (though physical
space cannot be filled up with spheres). This functional form is actually guided by the
reconstructed density profile obtained when removing the divergence by hand, as discussed in
\citesec{sec:reg_div_phase_space}. 

This prescription needs slight modifications when dealing with anisotropic systems since the
diverging term takes a different form in that case. In the constant-$\beta$ case this term is
proportional to $(\mathrm{d}\chi/\mathrm{d}\Psi)_{\Psi=0}$ where $\chi=r^{2\beta_{0}}\rho$, therefore
we propose the following profile:
\ben
\tilde{\rho} = \rho-\frac{\Psi_{\rm D}}{r^{2\beta_{0}}}
\left(\frac{\mathrm{d}^{n}\chi}{\mathrm{d}\Psi_{\rm D}^{n}}\right)_{\Psi_{\rm D}=0}\,,
\label{eq:alt_density_beta}
\een
where $\chi$ is defined in \citeeq{eq:density_beta} and $n$ is given in \citeeq{eq:index_beta}.
For the Osipkov-Merritt models
\ben
\tilde{\rho} = \rho - \frac{\Psi_{\rm D}}{1+r^{2}/r_{\rm a}^{2}}\left(\frac{\mathrm{d}\rho_{\rm OM}}{\mathrm{d}\Psi_{\rm D}}\right)_{\Psi_{\rm D}=0}\,.
\label{eq:alt_density_om}
\een
In these two cases, the gravitational potential is consistently calculated from
\citeeq{eq:alt_potential}. Note that in the anisotropic case, the modified profile
depends on the anisotropy variable ($\beta_{0}$ or $r_{\rm a}$) specific to the model.

The modified density and mass are compared to the original ones in
\citefig{fig:reconstructed_rho}.
The modified density differs from the original one in the outer part of the halo, and
underestimates the original profile by up to 40\% at $\sim 0.8\,R_{\rm max}$ (30\% at $R_{200}$)
in the isotropic case, which translates into a difference in the mass of only 15\% at
$r=R_{\rm max}$ (10\% at $R_{200}$). The inner, dynamically constrained part of the profile is
therefore kept
mostly unchanged by the prescription when the system is isotropic. The introduction of a
constant anisotropy causes a departure from the isotropic result in a systematic way that
depends on the sign of $\beta_{0}$. The difference in density and mass is higher in the
$\beta_{0}>0$ case than in the isotropic and $\beta_{0}<0$ cases, which is consistent with the
expectation for more radial orbits. The difference for Osipkov-Merritt models is even bigger:
the mass difference reaches 50\% at $R_{200}$. These differences can be understood in terms of the
anisotropy of the system. Our prescription removes matter at the edge of the halo to flatten the
density profile at $R_{\rm max}$. However, if the particles at $R_{\rm max}$ are mostly on radial
orbits, as is the case in the Osipkov-Merritt models where $\beta(R_{\rm max})\simeq 1$, they also
contribute to the density in the inner part of the halo. Therefore in a system with a high
positive $\beta$, removing matter in the outskirts also removes matter in the inner regions.
\begin{figure*}[!t]
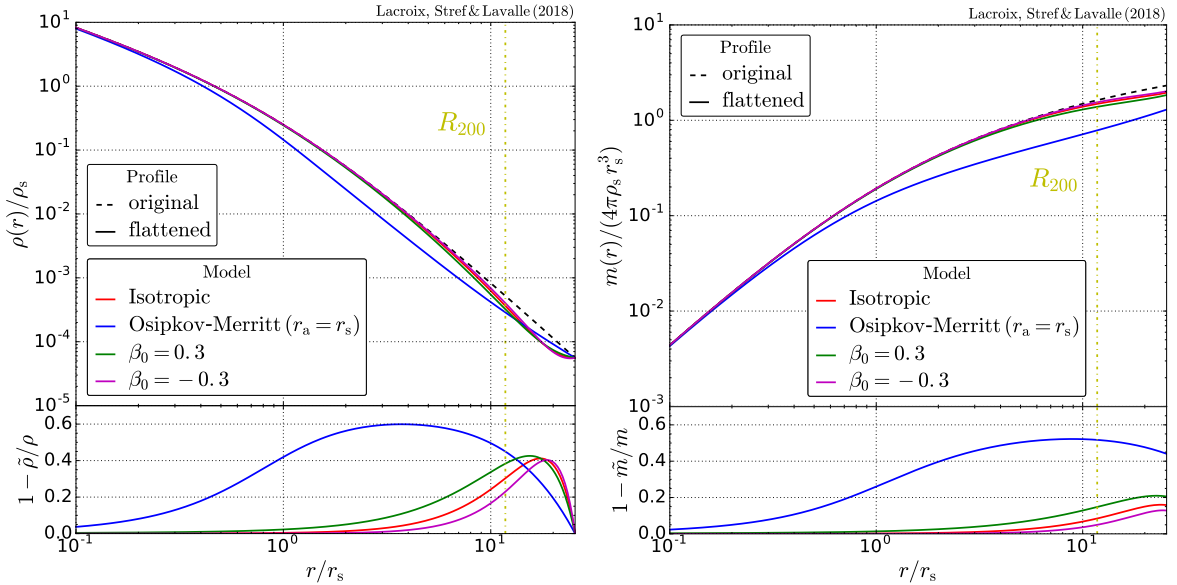

\centering
\includegraphics[width=0.495\linewidth]{{{fig_reconstructed_density}}}
\includegraphics[width=0.495\linewidth]{{{fig_reconstructed_mass}}}
\caption{\small \textbf{Left panel:} Density profile $\rho$ in the standard NFW case (black line)
  compared to the modified profiles for the isotropic, Osipkov-Merritt, $\beta = \pm 0.3$ (red,
  blue, green and magenta, respectively) defined in Eqs.~(\ref{eq:alt_density}),
  (\ref{eq:alt_density_beta}), (\ref{eq:alt_density_om}). The bottom panel shows the relative
  difference between the original and the modified profile. \textbf{Right panel:} Same as for left
  panel, for the mass profile.}
\label{fig:reconstructed_rho}
\end{figure*}

To summarize, we found that slight modifications of the density profile are enough
to get rid of the boundary-induced divergence in the isotropic case and in the case of
tangential anisotropy, while keeping the overall mass model consistent with the constrained
initial configuration. Indeed, the error induced on the Galactic mass at large radii is of order
$\sim 10\%$ in these cases. Therefore, such modifications preserve the self-consistency of the
formalism, and do not affect the density- and velocity-dependent observables related to
DM searches calculated in the inner $\sim$50-100 kpc of the MW. Errors of
$\gtrsim 10\%$ are expected in the outskirts, but should be negligible \eg\ when integrated
over the line of sight, like in the case of $p$-wave suppressed or Sommerfeld-enhanced
annihilation calculations. On the other hand, this regularization procedure fails in the
case of significant radial anisotropy, like in the Osipkov-Merritt model, unless the
anisotropy radius is taken very large (isotropy limit).
\subsubsection{Regularization through the phase-space distribution}
\label{sec:reg_div_phase_space}
The problem of the spatial boundaries in self-gravitating systems is rather classical
when one takes the DF as the fundamental characterizing function. A well-known example
is the so-called King model \cite{Woolley1954,WoolleyEtAl1956,King1962a,King1966,Michie1963,MichieEtAl1963}, meant to consistently describe bounded pseudo-isothermal systems (and applied to
globular clusters).
For the boundary-induced divergence at stake in our study, one can apply a similar procedure
by (i) cutting the non-physical diverging term in the phase-space DF $f({\cal E})$; (ii)
numerically deriving the modified gravitational potential from the Poisson equation fed by
the new DF
and appropriate boundary conditions (this is an important step which also defines the new mapping
between the potential and the radial coordinate); and (iii) integrating the new DF to get the
modified density profile. Although much more involved than the regularization through the
density presented in \citesec{sssec:reg_dens}, this method ensures to get a well-behaved
solution consistent with both the Boltzmann equation and the Poisson equation. It is
particularly well-suited to describe bounded systems like galaxies \cite{WidrowEtAl2005}, and
also to account for tidal effects induced by either neighboring systems like dwarf galaxies
\cite{DrakosEtAl2017,StrigariEtAl2017,PetacEtAl2018}, or hosted systems, like DM subhalos
\cite{BerezinskyEtAl2014,StrefEtAl2017}. In the present context, one still needs
then to make sure that the modified density profile does not depart too much from the initial
density profile, at least within the inner 50-100 kpc of the MW, not to spoil its
consistency with kinematic data.

Before inspecting possible ways of cutting the initial DF $f({\cal E})$, let us review
the full chain of calculations. Let us call $F(\tilde{\cal E})$ the modified DF after
truncation, where
\ben
\tilde{\cal E} = \tilde{\Psi} - \frac{\tilde{v}^2}{2}
\een
is the new energy associated with the system, $\tilde{\Psi}$ the new potential, and $\tilde{v}$
the new velocity coordinate. A priori, tilde quantities are different from non-tilde quantities
that pertain to the initial configuration. However, since $F(\tilde{\cal E})$ is known (inferred
from a modification of $f({\cal E})$ that we shall discuss later), we can fully determine the
DM component of the gravitational potential from the Poisson equation 
\ben
\Delta\tilde{\Psi}_{\rm D} &=& -4\pi G_{\rm N}\,\tilde{\rho}(\tilde{\Psi})
= -4\pi G_{\rm N}\int {\rm d}^3\vec{\tilde{v}}\,F(\tilde{\cal E})\nn\\
  &=& -4\pi G_{\rm N}\,
\left[\tilde{\rho}_0+
  4\pi\sqrt{2}\int_{0}^{\tilde{\Psi}}\sqrt{\tilde{\Psi}-\mathcal{E}}\,
  F(\mathcal{E})\,\mathrm{d}\mathcal{E}\right]\,,
\label{eq:poisson_tilde}
\een
where though the density profile $\tilde{\rho}$ is still undetermined, it is accessed through
the integral of the DF over the potential. Note that $\tilde{\Psi}=\tilde{\Psi}_{\rm D}+
\tilde{\Psi}_{\rm B}$, and that only the DM component is modified, such that
we actually take $\tilde{\Psi}_{\rm B}=\Psi_{\rm B}$. An important point here is that the mapping
between the radial coordinate and $\tilde{\Psi}$ is only defined through the Laplacian operator
$\Delta$ on the left-hand side, not on the right-hand side. Therefore, one needs appropriate
boundary conditions to solve this equation consistently with the physical system at hand.
In the present context, we are in principle forced to demand that $\tilde{\Psi}(R_{\rm max})=0$,
and since we do not want a significant departure from the initial potential in the inner parts of
the Galaxy, we further impose that $\mathrm{d}\tilde{\Psi}/\mathrm{d}r(0) =
\mathrm{d}\Psi/\mathrm{d}r(0)$. Besides, note that we allow for the presence of a constant
$\tilde{\rho}_0$ in the above equation, which is a free parameter and cannot be
recovered from the equation itself. This freedom in choosing the value of the density at the
boundary of the system is inherent to the Eddington formalism as previously seen in
\citeeq{eq:rho_reconstruction}---note that it can be neglected here as the density profile is no
longer an input in the regularization procedure, but an output. Finally, we stress that the above
differential equation has to be solved numerically.

We now discuss some possible forms for the modified DF $F$, which are to be considered as
ans\"atze aimed at recovering the non-diverging part of the initial DF while ensuring that
$F(\tilde{\cal E}\to 0)\to 0$.
We first consider the Eddington DF computed for a finite system with radial extension
$R_{\rm max}$. The initial DF is given in \citeeq{eq:eddington_form2} and diverges as
$\mathcal{E}\to0$. One way of modifying that DF to get a well-behaved distribution is simply to
remove the diverging term $\propto 1/\sqrt{\mathcal{E}}$. The modified DF is then
\ben
F(\tilde{\cal E}) = f(\tilde{\cal E})-\frac{1}{\sqrt{8}\pi^{2}}
\frac{1}{\sqrt{\tilde{\cal E}}}\left(\frac{\mathrm{d}\rho}{\mathrm{d}\Psi}\right)_{\Psi=0}
= \dfrac{1}{\sqrt{8}\pi^{2}} \int_{0}^{\tilde{\cal E}} \!
\dfrac{\mathrm{d}^{2}\rho}{\mathrm{d}\Psi^{2}} \,
\dfrac{\mathrm{d}\Psi}{\sqrt{\tilde{\cal E} - \Psi}}\,.
\label{eq:df_div_removed}
\een
Such an ansatz makes sense only if $\tilde{\cal E}$ spans the same range as ${\cal E}$ (note
that $\rho$ and $\Psi$ are non-tilde quantities). This is possible only if the condition
$\tilde{\Psi}_{\rm max}=\tilde{\Psi}(r=0)=\Psi(r=0)$ is obeyed, which is in contradiction with
the presumed boundary condition to solve \citeeq{eq:poisson_tilde}, \ie\
$\tilde{\Psi}(R_{\rm max})=0$. The latter condition must therefore be traded for the former in that
case, and the spatial boundary of the system is no longer $R_{\rm max}$, but a new
$\tilde{R}_{\rm max}$. In practice, though, we find that $\tilde{R}_{\rm max} \approx R_{\rm max}$,
such that the above ansatz can still be applied.

Since in the initial DF the diverging term becomes important only as $\mathcal{E}\to 0$, we expect
the modified potential to remain close to the original one which allows us to estimate the modified
density from the Abel equation
\ben
\frac{\mathrm{d}\tilde{\rho}}{\mathrm{d}\tilde{\Psi}} =
\sqrt{8}\pi\int_{0}^{\tilde{\Psi}}\frac{F(\mathcal{E})}{\sqrt{\tilde{\Psi}-\mathcal{E}}}\,
\mathrm{d}\mathcal{E}\,.
\label{eq:abel2}
\een
Assuming $\tilde{\Psi}\approx\Psi$ and $\tilde{\rho}(\tilde{\Psi}=0)\approx\rho(\Psi=0)$, we get
\ben
 \tilde{\rho} \approx \rho-\Psi\left(\frac{\mathrm{d}\rho}{\mathrm{d}\Psi}\right)_{\Psi=0}\,.
\een
The modification of the constant-$\beta$ DF is very similar, except the modification is only
performed on the energy-dependent part of the DF $G(\mathcal{E})$ in \citeeq{eq:df_constant_beta}.
The modification of the Osipkov-Merritt models is identical to the isotropic case with the change
$\mathcal{E}\rightarrow Q$. However, \citefig{fig:divergence_vesc_OM} shows that removing the
divergence by hand leads to a huge modification of the speed distribution. As a result, the
Osipkov-Merritt DF is very hard to regularize in a self-consistent way.

Note that the above expression for $\tilde{\rho}$ is similar to the one proposed in
\citeeq{eq:alt_density}, except that the potential that appears is the
DM only potential rather than the total potential. The density and mass shown in
\citefig{fig:reconstructed_rho} are therefore also relevant for the modified DF discussed here.

We now turn to a truncation of the DF more fundamentally inspired from the King model
\cite{King1966}. The original approach focused on making isothermal spheres finite in phase
space but was later generalized to generic mass distributions (see \eg\ \cite{WidrowEtAl2005}).
It was also very recently used to implement a realistic tidal truncation of satellite DM
halos \cite{DrakosEtAl2017}. The spirit of the method is slightly different from what was
presented just above in the sense that we no longer start from a diverging and ill-defined DF,
but from a well-behaved DF describing a self-gravitating system with spatial boundaries sent to
infinity (thereby resembling the King model, which starts from the Maxwellian DF that describes
the infinite isothermal sphere).
In the isotropic case, this initial DF is precisely the Eddington function $f({\cal E})$ given
in \citeeq{eq:eddington_form2}, taking the gravitational potential $\Psi(r)=-\phi(r)$ as the
solution to Poisson's equation with the boundary condition $\phi(r\to \infty)\to 0$.
We then implement a truncation in energy related to the desired radial boundary $R_{\rm max}$
from a procedure similar to the one introduced above: (i) cut the phase-space volume in energy
below a cutoff ${\cal E}_c=\Psi_0=\Psi(R_{\rm max})$; (ii) define a new phase-space DF
$F(\tilde{\cal E}\equiv {\cal E}-\Psi_0)$ from $f({\cal E})$ above the cutoff, with the expected
asymptotic behavior $F(\tilde{\cal E}\to 0)\to 0$; (iii) determine the new associated gravitational
potential $\tilde{\Psi}$ from \citeeq{eq:poisson_tilde} (as previously, this defines the new
mapping between the radial coordinate and the potential); (iv) integrate the new DF to get
the modified density profile $\tilde{\rho}$.

According to this procedure, the ansatz for the modified DF $F$ that relates a cutoff in energy
to a radial cutoff is then defined as
\ben
F(\tilde{\cal E}) = \left\{
\begin{array}{ll}
  f(\tilde{\cal E}+\Psi_{0})-f(\Psi_{0})\, &\mathrm{for}\,\tilde{\cal E}\geqslant 0\\
  0\, &\mathrm{for}\,\tilde{\cal E}<0
\end{array}
\right.
\label{eq:king_model}
\een
This DF is continuous and satisfies $F(\tilde{\cal E}=0)=0$ by construction. The associated
gravitational potential $\tilde{\Psi}$ is solution of the Poisson equation
\citeeq{eq:poisson_tilde}, with initial conditions to be specified. If we set the cutoff in the
initial DF to ${\cal E}_c=\Psi_0=\Psi(R_{\rm max})$, then
$\tilde{\Psi}_{\rm max}=\Psi_{\rm max}-\Psi(R_{\rm max})$ by construction, which by no means
guarantees that $\tilde{\Psi}$ vanishes at $R_{\rm max}$. In practice though, we find that the
radius $\tilde{R}_{\rm max}$ at which $\tilde{\Psi}(\tilde{R}_{\rm max})=0$ is very close to
$R_{\rm max}$, though slightly larger. To get $\tilde{\Psi}(R_{\rm max})=0$ directly from the
Poisson equation, one would instead need to tune the initial cutoff potential $\Psi_0$
until equality is reached---in the same vein, we find in that case that $\Psi_0\approx
\Psi(R_{\rm max})$.

Note that unlike removing the diverging term
``by hand", the King approach may lead to a physical interpretation in terms of tidal cut,
since it has been shown in numerical simulations that tidal stripping tends to remove
particles based on heir energy rather than their angular momentum \cite{ChoiEtAl2009}. In the
present context, such stripping could have resulted from gravitational interactions with the
neighboring galaxies. We show the dark halo profile reconstructed from the DF of
\citeeq{eq:king_model} after solving \citeeq{eq:poisson_tilde} in
\citefig{fig:reconstructed_rho_king}, where the difference in setting the cutoff discussed
just above is illustrated explicitly.
\subsubsection{Regularization of the boundary-induced divergence: Summary}
\label{sssec:v_div_summ}
Here we summarize the pros and cons of the regularization procedures implemented above to remove the
radial boundary-induced divergence of the phase-space DF as ${\cal E}\to 0$ or equivalently
$v\to v_{\rm esc}$. For the isotropic case, we saw that the technically easiest way to remove the
divergence while ensuring the self-consistency of the Eddington inversion method was to slightly
modify the input density profile around the radial boundary $R_{\rm max}$ in such a way that the
dynamics is unaffected in the central regions of the Galaxy. In that case, one can
straightforwardly find the new gravitational potential $\tilde{\Psi}_{\rm D}$ by directly
integrating the Poisson equation over the radial coordinate from \citeeq{eq:relative_potential}.
The regularization through modifications of the DF is more involved as it requires to calculate
$\tilde{\Psi}_{\rm D}$ by numerically solving the Poisson equation. This is the only way to recover
a mapping between the potential and the radial coordinate, and then to compute the resulting
modified density profile. Both methods give similar distortions to the initial density profile,
which lie within the current statistical and systematic uncertainties on the dark halo mass
profile. Ultimately, the best, while much more involved approach, would be to start from
well-defined DF and profile before performing the likelihood analysis to account for the
kinematic constraints and to get best-fitting Galactic mass models, similar to the action-angle
analyses (\eg~\cite{BinneyEtAl2015}). This goes beyond the scope of this paper.

For anisotropic systems, we saw that both methods may apply to tangential anisotropy ($\beta<0$),
but fail for radial anisotropy (both $\beta>0$ and the Osipkov-Merritt models). In the latter
case, the only way to get finite results is to remove the diverging term
($\propto 1/\sqrt{\cal E}$ or $\propto 1/\sqrt{Q}$) by hand, but this is at the cost of a
meaningful and self-consistent normalization of the phase-space DF. Therefore, we are then left
with a theory that is no longer a self-consistent inversion of the integral
\citeeq{eq:rho_definition}, whose DF must be normalized to unity (or $\rho$ or $\tilde{\rho}$)
by hand and is no longer simply related to the DM density profile. Although such
a DF might be perfectly licit as a description of a gravitational system, its theoretical status
appears unclear to us.


\subsection{Positivity and stability issues}
\label{ssec:gamma} 
We now move to another kind of issues that may arise in the Eddington formalism:
the potential breakdown of the inversion, very often due to the presence of baryonic components.
More concretely, it turns out that some perfectly sound configurations of Galactic mass
models may lead to ill-defined DFs through this method, which are the manifestation of unstable
configurations in phase space. In these cases, Eddington-like inversions can no longer be used to
self-consistently describe the DM halo, because some degrees of freedom are likely
missing to make full physical sense of the DM component (axisymmetry, action-angles
coordinates, \etc.). We stress that the potential breakdown of the Eddington formalism may
only manifest itself in some regions of the phase space. This is actually barely checked in the
context of predictions for direct DM searches. A typical signature of such a breakdown
is a DF exhibiting negative values in specific regions of phase space, which will be
discussed in \citesec{sssec:pos}. More subtle while complementary considerations linked to the
stability of gravitational systems will be discussed in \citesec{ssec:gamma}.

\subsubsection{Positive distribution functions}
\label{sssec:pos}
A trivial requirement for a DF to be well-behaved is positivity everywhere,
\ie~$f(\vec{r},\vec{v})\geqslant0$ for any $(\vec{r},\vec{v})$.
Although most dark halo shapes are fully Eddington invertible for
DM-only systems (\eg~\cite{Widrow2000}), there is in general no guarantee that
Eddington's inversion leads to a DF positive all over the halo for any given pair of
DM density profile $\rho$ and total gravitational potential $\Psi=\Psi_{\rm D}+\Psi_{\rm B}$.
We will inspect below the specific case of cored profiles, but it usually turns out that the
presence of a baryonic component, which breaks the plain correlation between the density and
the potential, may drive the DF negative in some regions of the system.

Sufficient conditions for positivity were identified in Refs.~\cite{CiottiEtAl1992,Ciotti1996} for
the Osipkov-Merritt models, in the general case of multi-component systems. From
\citeeq{eq:om_form1} we can identify a necessary condition for the positivity of $f_{\rm OM}$,
which is
\ben
\frac{\mathrm{d}\rho_{\rm OM}}{\mathrm{d}\Psi}\geqslant
0~{\rm for}\,0\leqslant\Psi\leqslant\Psi_{\rm max}.
\een
In this equation, $\rho_{\rm OM}$ corresponds to the DM while $\Psi=\Psi_{\rm D}+\Psi_{\rm B}$ is the
total potential (from the DM plus baryons). If this necessary condition is satisfied, a sufficient
condition for positivity is \cite{CiottiEtAl1992,Ciotti1996} 
\ben
\frac{\mathrm{d}}{\mathrm{d}\Psi_{\rm D}}
\left[\frac{\mathrm{d}\rho_{\rm OM}}{\mathrm{d}\Psi_{\rm D}}
  \left(\frac{\mathrm{d}\Psi}{\mathrm{d}\Psi_{\rm D}}\right)^{-1}\sqrt{\Psi}\right]\geq 0
~\forall\,0\leqslant\Psi\leqslant\Psi_{\rm max}.
\een
One can readily see that these conditions are also valid for isotropic systems as well as single
component systems. All McM17 halo profiles verify this condition.

Let us return to the isotropic case and inspect it in detail. Most standard
single-component mass
distributions (\eg~NFW, Einasto, \etc.) have well-defined ergodic DFs \cite{Widrow2000}. Yet,
some well-motivated profiles do lead to a negative DF. Troublesome profiles can be identified
using \citeeq{eq:abel_equation_sec2}. If the derivative $\mathrm{d}\rho/\mathrm{d}\Psi_{\rm D}$
cancels for some values of $\Psi_{\rm D}$, then \citeeq{eq:abel_equation_sec2} forces $f$ to take
negative values. This is expected to happen if the DM profile if very flat somewhere, as is the
case for cored distributions for instance. In the case of single-component systems, the left-hand
side of \citeeq{eq:abel_equation_sec2} can be written
\ben
\frac{{\rm d}\rho}{{\rm d}\Psi_{\rm D}} =
\frac{{\rm d}r}{{\rm d}\Psi_{\rm D}}\frac{\mathrm{d}\rho}{\mathrm{d}r} =
-\frac{r^{2}}{G\,m_{\rm D}(r)}\frac{\mathrm{d}\rho}{\mathrm{d}r}\,,
\label{eq:drho_dpsi}
\een
where the mass $m_{\rm D}(r)$ is related to the density $\rho(r)$ through \citeeq{eq:mass}. Let us
now consider as an example the following class of cored DM density profiles:
\ben
\rho(r) = \rho_{\rm s}\left[1+\left(\frac{r}{r_{\rm s}}\right)^{\alpha}\right]^{-\beta/\alpha}\,,
\label{eq:alpha_beta_density}
\een
with $\alpha>0$ and $\beta>0$. In the limit $r\to 0$ (equivalently
$\Psi_{\rm D} \to \Psi_{\rm max}$) we have $\mathrm{d}{\rho}/\mathrm{d}r\propto r^{\alpha-1}$
and $m\propto r^{3}$, therefore $\mathrm{d}\rho/\mathrm{d}\Psi_{\rm D} \propto r^{\alpha-2}$. The
asymptotic value of the derivative is then non-zero only if $\alpha\leqslant 2$. Consequently,
for \textit{any} single-component system with a density profile given by
\citeeq{eq:alpha_beta_density} and with $\alpha>2$, the Eddington method leads to a negative
ergodic DF.
We stress that
\ben
0<\alpha\leqslant 2 \;\;\text{is a \textit{necessary} condition (isotropic case)}
\label{eq:cond_alpha_dm}
\een
to get a positive DF for a DM-only system. However, it is certainly not sufficient for a
multi-component system. Since the argument is based on the asymptotic behavior of
$\mathrm{d}\rho/\mathrm{d}\Psi_{\rm D}$ as $r\rightarrow0$, our result holds for Osipkov-Merritt
models as well since the associated anisotropy goes to zero when $r\ll r_{\rm a}$. In the constant
anisotropy case the situation is different because an artificial slope $2\beta_{0}$ is present in
the Abel equation given in \citeeq{eq:abel_eq_constant_beta}. Consequently, if the density profile
$\rho$ has an inner slope $-\gamma$, the pseudo-density $\chi$ has an inner slope
$2\beta_{0}-\gamma$. Note that a requirement for Eddington's method and its extensions to work
is that the generalized density ($\rho,\ \rho_{\rm OM},\ \chi$ depending on the model) is a growing
function of $\Psi$. Therefore,
\ben
2\beta_{0}\leq \gamma\;\;\text{is a \textit{necessary} condition (anisotropic $\beta_0$ case)}
\label{eq:cond_beta0}
\een
to get a positive constant-anisotropy DF.
This forbids for instance any cored system to have a constant, positive anisotropy, and in
general sets an upper limit on the constant anisotropy a system can feature. This is
a subset of a more general slope-anisotropy inequality \cite{AnEtAl2006,CiottiEtAl2010}.

Adding a baryonic component to the system can affect these results. If the DM profile follows
\citeeq{eq:alpha_beta_density} and the baryonic profile is cored, the low-radius behavior of
$\mathrm{d}\rho/\mathrm{d}\Psi$ (with $\Psi=\Psi_{\rm D}+\Psi_{\rm B}$ the total potential) is
unchanged with respect to that of $\mathrm{d}\rho/\mathrm{d}\Psi_{\rm D}$. Therefore, the positivity
condition remains $\alpha\leqslant 2$. If the baryonic density profile is cuspy with inner slope
$-\gamma_{\rm B}$ (\eg\, $\gamma_{\rm B}=1$ for a Hernquist profile), the result is modified. The
mass is now dominated by the baryonic component as $r\rightarrow0$, and we have
$\mathrm{d}\rho/\mathrm{d}\Psi\propto r^{\alpha-2+\gamma_{\rm B}}$. The necessary condition for
positivity becomes
\ben
0<\alpha\leqslant 2-\gamma_{\rm B}\,
\label{eq:cond_alpha_bar}
\een
\ie\ baryons reduce the parameter space providing a positive DF.

\subsubsection{Stable distribution functions}
\label{sssec:stab}
%
We would like to stress here that positivity is not strong enough a criterion for a DF to
give a satisfactory description of a DM halo. Indeed, some $(\rho,\Psi)$ pairs satisfying the
positivity conditions can still lead to a DF that is an \textit{unstable} solution of the
collisionless Boltzmann equation. Some conditions for stability against different kinds of
perturbations are reviewed in Ref.~\cite{BinneyEtAl2008}. A result of interest for us is
Antonov's second law \cite{Antonov1962,Lebovitz1965,Lynden-BellEtAl1968} which guarantees the
stability of an ergodic DF $f$ against non-radial modes if $\mathrm{d}f/\mathrm{d}\mathcal{E}>0$.
A complementary result is the Doremus-Feix-Baumann theorem \cite{DoremusEtAl1971,KandrupEtAl1985},
which ensures stability against radial modes if $\mathrm{d}f/\mathrm{d}\mathcal{E}>0$.
Consequently, a \textit{sufficient condition for the stability} of ergodic DFs $f(\mathcal{E})$
against all perturbations is
\ben
\frac{\mathrm{d}f}{\mathrm{d}\mathcal{E}}(\mathcal{E})>0~{\rm for\,all}\,\mathcal{E}\,.
\label{eq:stability1}
\een
We now investigate the consequences of this condition on DM density profiles. In practice
we use profiles of the form of \citeeq{eq:alt_density} in order to get rid of the divergence
discussed in \citesec{ssec:rmax}. Note that we previously established that this divergence is a
sign of an artificial compression of the phase space, but it can also be viewed as an
unstable configuration as it violates the stability
criterion given in \citeeq{eq:stability1}. We wish to find a more convenient criterion involving
the density profile and the potential rather than the DF itself. We recall the expression of the
DF when the boundary term is zero:
\ben
f(\mathcal{E}) = \int_{0}^{\mathcal{E}}\frac{\mathrm{d}^{2}\rho}{\mathrm{d}\Psi^{2}}
\frac{1}{\sqrt{\mathcal{E}-\Psi}}\,\mathrm{d}\Psi\,.
\label{eq:df_without_divergence}
\een
From this expression we see that $\mathrm{d}f/\mathrm{d}\mathcal{E}>0,\,\forall\mathcal{E}$ only
if $\mathrm{d}^{2}\rho/\mathrm{d}\Psi^{2}>0,\,\forall\Psi$. Moreover, starting from the Abel
equation in \citeeq{eq:abel_equation_sec2} and performing an integration by parts, we get
\ben
\dfrac{\mathrm{d}\rho}{\mathrm{d}\Psi} = 2\sqrt{8}\pi\int_{0}^{\Psi}
\sqrt{\Psi-\mathcal{E}}\,\frac{\mathrm{d}f}{\mathrm{d}\mathcal{E}}\,\mathrm{d}\mathcal{E}\,,
\een
which implies that $\mathrm{d}^{2}\rho/\mathrm{d}\Psi^{2}>0,\,\forall\Psi$ only if
$\mathrm{d}f/\mathrm{d}\mathcal{E}>0,\,\forall\mathcal{E}$. To summarize, we have
\ben
\frac{\mathrm{d}f}{\mathrm{d}\mathcal{E}}>0,~\forall\mathcal{E} \iff
\frac{\mathrm{d}^{2}\rho}{\mathrm{d}\Psi^{2}}>0,~\forall\Psi\,.
\label{eq:stability2}
\een
Therefore, the stability criterion takes the very simple following form:
$\mathrm{d}^{2}\rho/\mathrm{d}\Psi^{2}>0,\,\forall\Psi$. From \citeeq{eq:df_without_divergence}, it
is obvious that this criterion is also a sufficient condition for positivity. From now on, we
consider \citeeq{eq:stability2} as defining the range of applicability of the Eddington formalism
since a system that violates this condition could lead to unstable phase-space
configurations or a negative DF.

The stability criterion of \citeeq{eq:stability2} can be extended in part to non-isotropic
spherical systems with a DF of the form $f(\mathcal{E},L)$. It is shown in
Ref.~\cite{DoremusEtAl1973} that
systems satisfying $\partial f/\partial\mathcal{E}>0$ for all $(\mathcal{E},L)$ are stable against
radial perturbations. This is directly applicable to the constant-$\beta$
(Eq.~\ref{eq:df_constant_beta}) and Osipkov-Merritt (Eq.~\ref{eq:df_om}) models, resulting in
\begin{subequations}
  \label{eq:stability_anisotropy}
  \ben
  \frac{\mathrm{d} G}{\mathrm{d}\mathcal{E}}&>&0,~\forall \mathcal{E}\\
  \frac{\mathrm{d} f_{\rm OM}}{\mathrm{d}Q}&>&0,~\forall Q\,.
  \een
\end{subequations}
However, the response of anisotropic systems to non-radial perturbations is much more complex,
due to the possibility of radial-orbit instabilities, so that no simple stability criteria are
known. Analytical studies are usually involved
(\eg~\cite{Antonov1987,PerezEtAl1996,ReinEtAl2003,MarechalEtAl2010}), and the stability
properties of anisotropic systems are very often investigated thanks to numerical simulations
(\eg~\cite{MerrittEtAl1985,BarnesEtAl1986,MezaEtAl1997}), which is far beyond the scope
of this work. In the following, we rely on the criterion given in
\citeeq{eq:stability_anisotropy}, which should be understood as necessary rather than sufficient.

\begin{figure*}[t!]
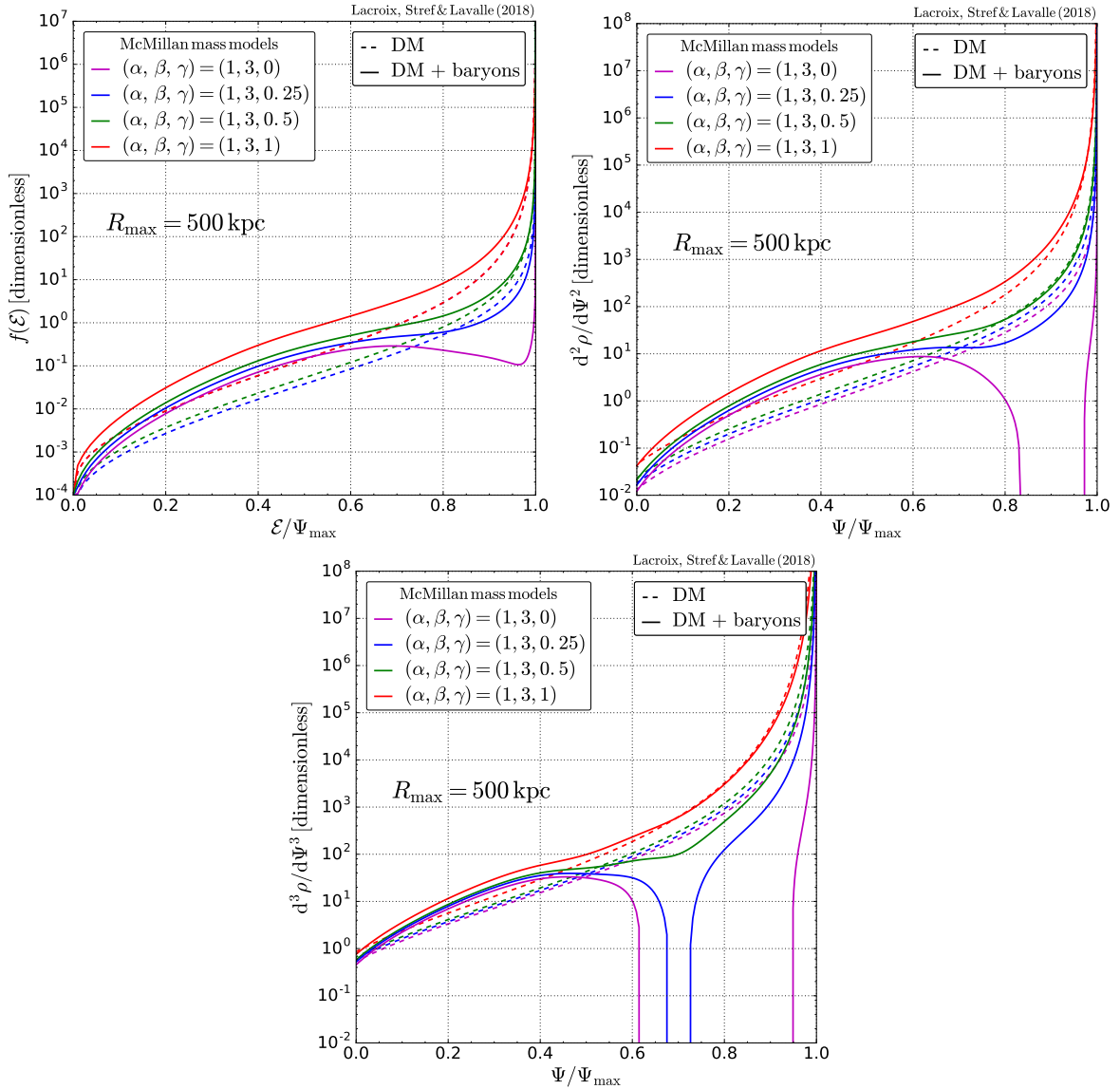

\centering
\includegraphics[width = 0.495\linewidth]{{{fig_df_mcmillan_models}}}
\includegraphics[width = 0.495\linewidth]{{{fig_d2rhodPsi2_mcmillan}}}
\includegraphics[width = 0.495\linewidth]{{{fig_d3rhodPsi3_mcmillan}}}
\caption{\small \textbf{Top left panel}: Ergodic distribution functions for several mass models from
  Ref.~\cite{McMillan2017}. The DFs are in units of
  $(4\pi G_{\rm N})^{-3/2}\rho_{\rm s}^{-1/2}r_{\rm s}^{-3}$. \textbf{Top right panel}:
  Second derivative of the density $\rho$ with respect to the total potential $\Psi$. The
  derivative is in units of $(4\pi G_{\rm N})^{-2}\rho_{\rm s}^{-1}r_{\rm s}^{-4}$.
  {\bf Bottom panel}: Third derivative ${\rm d}^3\rho/d\Psi^3$ in units of
  $(4\pi G_{\rm N})^{-3}\rho_{\rm s}^{-3/2}r_{\rm s}^{-6}$. }
\label{fig:df_mcmillan_models}
\end{figure*}

We investigated the stability of the phase-space configurations obtained by
Eddington-inverting realistic and kinematically constrained McM17 MW dark halos
\cite{McMillan2017}. Shown in the top left panel of \citefig{fig:df_mcmillan_models} are the
isotropic DFs for each mass model, both with and without the baryonic contribution to the
potential $\Psi$. To simplify the discussion, the DFs are shown \textit{without} the
diverging term discussed in \citesec{ssec:rmax}, and without any regularization plugged in.
Indeed, we will see that in these examples, instabilities manifest themselves mostly in the
central regions of the Galaxy, \ie~${\cal E}/\Psi_{\rm max}\gtrsim 0.5$. The dark halos shown
in the figure mostly differ in the inner slope $\gamma$ of the density profile. In the
absence of baryons (dashed lines), all the DFs satisfy the stability criterion given in
\citeeq{eq:stability1} and are therefore stable. We explicitly verified that the models also
satisfy the condition in \citeeq{eq:stability2} by plotting the second-order derivative
$\mathrm{d}^{2}\rho/\mathrm{d}\Psi^{2}$ in the top right panel of \citefig{fig:df_mcmillan_models}.

The situation changes when baryons are added to the potential. Then the DFs flatten at high
energy (toward the central regions), and may even turn into a dip, as is the case of the DM core
($\gamma=0$, solid magenta line), which violates the stability criterion in
\citeeq{eq:stability1}. The derivative
$\mathrm{d}^{2}\rho/\mathrm{d}\Psi^{2}$ takes negative values in that case and the stability
criterion in \citeeq{eq:stability2} is also violated as expected. This mass model is therefore
very likely to correspond to an unstable phase-space configuration.\footnote{More precisely,
  the initial assumption of ergodicity cannot accommodate this density-potential pair; one
  would need to increase the number of degrees of freedom in phase space to find a stable DF.}
The presence of a dip in the ergodic DF has direct consequences in the speed distribution
defined in \citeeq{eq:v_df}. In the left panel of \citefig{fig:fv_mass_ratio}, we
show the speed distributions for the different mass models at $r=0.01\,\rm kpc$,
\ie~corresponding to regions where the energy range probes the dip. The speed distribution
of the unstable model (magenta line) exhibits a very strong double-peak feature: a very large peak
at $v\sim 450\,\rm km/s$, and a much smaller one at $v\sim 50\,\rm km/s$.
The phase-space distribution is somewhat artificially forced to large velocities to allow for a
kinetic pressure strong enough to prevent the halo from collapsing to a cusp---hardly a stable
configuration in the isotropic case. The appearance of such a double-peak feature is
characteristic of a troublesome configuration, and we stress that it has to be checked all over
the halo (equivalently all over the available energy range). Indeed, the same problematic model
would have given a perfectly licit speed distribution at larger radii (where the energy range
would not probe the dip in the DF). We note, however, that it is not straightforward to firmly
analyze this feature in terms of instability since it is also present in the $\gamma=0.25$ case,
which satisfies the stability criterion, while clearly exhibiting a transition to a double-peak
distribution. This can be seen in the blue curve of \citefig{fig:fv_mass_ratio}. In fact, as can
readily be guessed from \citeeq{eq:eddington_form2} (the part in brackets) and from both the top
right and the bottom panels of \citefig{fig:df_mcmillan_models}, a way to select
better-behaved speed distributions (without double-peak feature) is simply to impose an additional
criterion based on the third derivative instead of the second:
\ben
\frac{\mathrm{d}^{3}\rho}{\mathrm{d}\Psi^{3}}>0,~\forall\Psi\,.
\label{eq:stability3}
\een
In the following, we will remain agnostic about the origin of this two-peak behavior and just
stick to the stability criterion of \citeeq{eq:stability2}, keeping in mind that
\citeeq{eq:stability3} could further be applied to remove controversial cases. We therefore keep
the McM17 $\gamma=0.25$ case as viable, while we reject the $\gamma=0$ case.

We now wish to characterize in more detail the instability when baryons contribute to the
potential. We write the mass of the system as $m=m_{\rm D}+m_{\rm B}$ and the gravitational
potential as $\Psi=\Psi_{\rm D}+\Psi_{\rm B}$. Then the derivative that appears in the stability
criterion can be written
\ben
\frac{\mathrm{d}^{2}\rho}{\mathrm{d}\Psi^{2}} =
\left(\frac{m_{\rm D}}{m_{\rm D}+m_{\rm B}}\right)^{2}
\left[\frac{\mathrm{d}^{2}\rho}{\mathrm{d}\Psi_{\rm D}^{2}}-
  \frac{\mathrm{d}\rho}{\mathrm{d}\Psi_{\rm D}}
  \frac{\mathrm{d}}{\mathrm{d}\Psi}\left(\frac{m_{\rm B}}{m_{\rm D}}\right)\right].
\een
From this expression, we get the sufficient condition for stability:
\ben
\frac{\mathrm{d}^{2}\rho}{\mathrm{d}\Psi_{\rm D}^{2}}
\left/\frac{\mathrm{d}\rho}{\mathrm{d}\Psi_{\rm D}}\right. >
\frac{\mathrm{d}}{\mathrm{d}\Psi}\left(\frac{m_{\rm B}}{m_{\rm D}}\right)\,.
\label{eq:stability_baryons}
\een
The quantities appearing on the left-hand side of \citeeq{eq:stability_baryons} only refer to
DM, while baryons appear on the right-hand side through their mass $m_{\rm B}=m_{\rm B}(r)$ and the
total potential $\Psi$. In the absence of baryons, $m_{\rm B}=0$ and \citeeq{eq:stability_baryons}
simplifies to $\mathrm{d}^{2}\rho/\mathrm{d}\Psi_{\rm D}^{2}>0$ which is exactly the stability
criterion in the DM-only case. Let us discuss the right-hand term in more detail. The
baryonic mass is present in the ratio $m_{\rm B}/m_{\rm D}$ and in the potential $\Psi$, so we do
not expect it to be the most important parameter here. Rather, the spatial extension of the
baryonic distribution with respect to the DM one is the relevant factor. To illustrate this,
we show the ratio $m_{\rm B}/m_{\rm D}$ as a function of $\Psi$ in the right panel of
\citefig{fig:fv_mass_ratio}. We show the isolated contribution of the bulge and the disk, as well
as the total baryonic contribution. We can see that the bulge-to-DM ratio is steeper than both
the disk-to-DM and baryons-to-DM ratios. The bulge-only configuration is therefore more likely
to be inconsistent with the ergodic assumption than the disk-only configuration and the
full mass model, even though the baryonic mass is much less important in that case.

\begin{figure*}[!t]
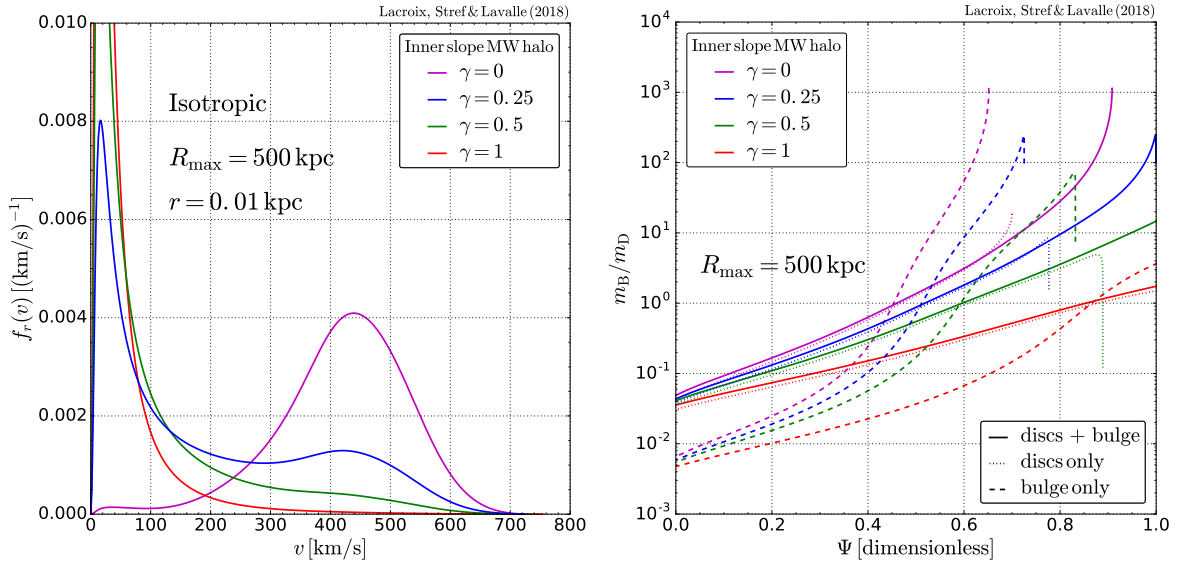

  \centering
  \includegraphics[width=0.502\linewidth]{{{fig_fv_r1e-2kpc}}}
  \includegraphics[width=0.488\linewidth]{{{fig_mass_ratio_mcmillan}}}
  \caption{\small \textbf{Left panel}: Speed distribution at $r=0.01\,\rm kpc$ for the mass models
    of \cite{McMillan2017}, computed from the ergodic DF in \citeeq{eq:df_without_divergence}.
    \textbf{Right panel}: Ratio of the baryonic mass to the DM mass as a function of
    the total potential $\Psi$ (in units of $4\pi G_{\rm N}\rho_{\rm s}r_{\rm s}^{2}$).}
  \label{fig:fv_mass_ratio}
\end{figure*}

We also investigated the effects on the second derivative $\mathrm{d}^{2}\rho/\mathrm{d}\Psi^{2}$
of changing the bulge characteristic mass density $\rho_{\rm 0,b}$ and radius $r_{\rm b}$
(see Eq.~\ref{eq:bulge}), while keeping the disks parameters fixed. Results are shown in
\citefig{fig:contours}, where we plot
$r_{\rm b}/r_{\rm s}$ as a function of $\rho_{\rm 0,b}/\rho_{\rm s}$. The bulge parameters are scaled
to the dark halo parameters. The points correspond to the McM17 mass models \cite{McMillan2017}
for $\gamma=0$, 0.25, 0.5. Those three models have nearly identical values for the bulge
parameters, the difference in coordinates only comes from the change in the halo parameters
$\rho_{\rm s}$ and $r_{\rm s}$. The red shaded areas are the portions of parameter space where
$\mathrm{d}^{2}\rho/\mathrm{d}\Psi^{2}$ goes negative, \ie~the Eddington DF violates the stability
criterion. We can see in \citefig{fig:contours} that the $\gamma=0$ mass model point is inside
the $\gamma=0$ excluded area, while the $\gamma=0.25$ and $\gamma=0.5$ models are
in their allowed regions. This is in agreement with the right panel of
\citefig{fig:df_mcmillan_models}, where the $\gamma=0$ case is explicitly shown to violate
the stability criterion. This figure further allows one to easily check whether one's favorite
Galactic mass model can be Eddington inverted.

\begin{figure}[!t]
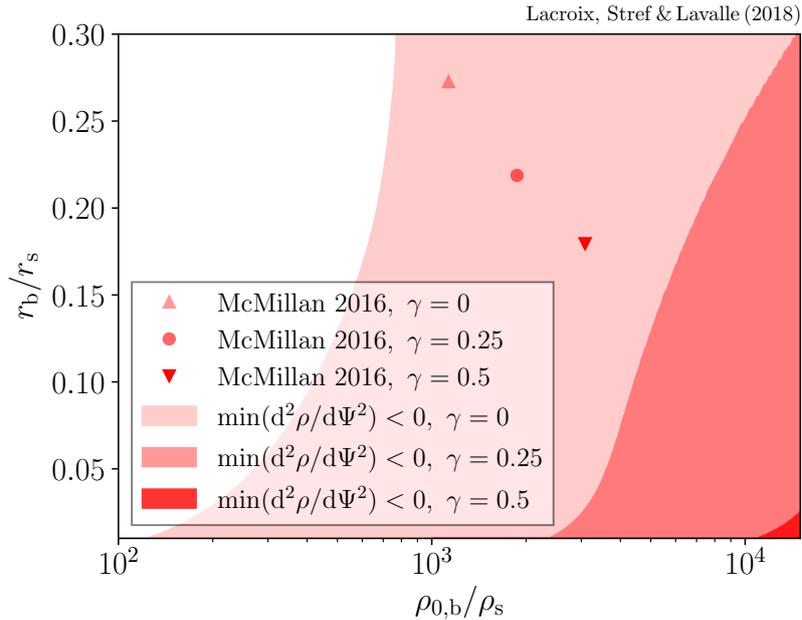

\centering
\includegraphics[width=0.7\linewidth]{{{fig_contours_McMillan_bulge_plus_disk}}}
\caption{\small Sign of the minimum of $\mathrm{d}^{2}\rho/\mathrm{d}\Psi^{2}$ on the plane
  ($\rho_{\rm 0,b}/\rho_{\rm s},r_{\rm b}/r_{\rm s}$) with $\rho_{\rm 0,b}$ ($\rho_{\rm s}$) and $r_{\rm b}$
  ($r_{\rm s}$) the characteristic density and radius of the bulge (dark halo). The parameters of the disk are fixed to the MCM17 values. Results are shown
  for $\gamma=0, 0.25,0.5$ with $\gamma$ the inner slope of the DM profile. Points
  indicate the positions of the McM17 models.}
\label{fig:contours}
\end{figure}

\subsubsection{Positive and stable distribution functions: summary}
\label{sssec:stab_summ}

In this section, we have discussed several theoretical issues that arise when
trying to infer the DF of a galactic system in a self-consistent way with the Eddington
formalism and its most simple anisotropic extensions. We have established its validity range,
and provided prescriptions to deal with these issues. These prescriptions can be readily
used to ensure a self-consistent application of Eddington-like inversions.

We have first discussed in \citesec{sssec:pos} the conditions to get a DF positive over the whole
energy range---this is mostly relevant to systems with both DM and baryons. For a DM
profile of the $\alpha\beta\gamma$ type [see \citeeq{eq:halo}], we isolated a rather simple
necessary condition on the index $\alpha$ given in \citeeq{eq:cond_alpha_bar}, which forces the
transition between the asymptotic indices $\gamma$ and $\beta$ to be smoother and smoother as
the baryonic distribution steepens in the very central parts of the Galaxy---this also applies
to the anisotropic Osipkov-Merritt model, which actually tends to the isotropic case when
$r\ll r_{\rm a}$. For constant-anisotropy models, a necessary condition exists in terms of
$\beta_0$, given in \citeeq{eq:cond_beta0}.
  
We have then discussed in \citesec{sssec:stab} more fundamental features which can be related
to the (in)stability of gravitational systems. We have shown that the Eddington inversion
can only provide a well-behaved DF when the condition given in \citeeq{eq:stability2} is
fulfilled. An even more stringent condition providing an unambiguous speed distribution
is given in \citeeq{eq:stability3}. In contrast to the positivity issue though, stability
conditions cannot be derived in the anisotropic cases, except for the very special case of
radial perturbations. In that case, the stability conditions are given in
\citeeq{eq:stability_anisotropy}.

Finally, we showed in \citefig{fig:contours} how to quickly check whether realistic Galactic
mass models are Eddington-invertible, only from the bulge-to-halo ratio of the scale densities.
This figure can be used as a preliminary diagnosis before going into more involved calculations.
In any case, all the discussion developed in this section fully applies to the general case, for
systems with or without baryons.

\section{Impact on predictions for dark matter searches}
\label{sec:dd}
In this section, we study the impact of the issues discussed in the previous sections on
predictions for DM searches. We shall obviously focus on velocity-dependent observables, and more
particularly on observables related to both direct DM searches and indirect DM searches:
the moments (and inverse moments) of the DM speed (relevant to direct DM searches, DM capture
by stars, or PBH microlensing), and the moments (an inverse moments) of the two-DM-particle
relative speed (relevant to $p$-wave-suppressed or Sommerfeld-enhanced DM annihilation).
We will embed the former observables in a {\em direct-search} class, while the latter
will define the {\em indirect-search} class, to make more explicit contact with the WIMP
phenomenology. We shall make quantitative comparisons between the self-consistent Eddington
approach (whenever applicable) and the Maxwellian approximation, which is most commonly used in
this context.
Note that the Maxwell-Boltzmann (MB) DF or velocity distribution is consistent with the
collisionless Boltzmann equation only if the underlying density profile is an infinite isothermal
sphere that also dominates the potential. It is therefore by no means theoretically consistent
with the input dark halo profile we will consider in the following calculations, but is usually
assumed to provide a ``reasonable'' approximation. Like in the isothermal sphere, we will
still use the link between the 3D velocity dispersion $\sigma$ and the circular velocity as
follows:
\ben
\sqrt{\frac{2}{3}}\,\sigma = v_{\rm circ}(r) = \sqrt{\frac{G_{\rm N}\,m(r)}{r}}\,,
\een
with $m(r)$ consistently derived from the mass model. Therefore, the dispersion velocity
associated with the MB DF will be radial dependent in the following.

\subsection{Direct-search-like observables}
Let us define a generic function for the moments of the DM speed in the Galactic frame:
\begin{subequations}
  \ben
  \Xi_n(v_{\rm min},v_{\rm max},r) &\equiv&\omega^{-1}(r)
  \int_{v_{\rm min} \leqslant |\vec{v}| \leqslant v_{\rm max}}
      {\rm d}^{3} \vec{v} \, |\vec{v}|^n \, f_{\vec{v}}(\vec{v},r)\\
      \omega(r) &\equiv &\int {\rm d}^{3} \vec{v} \, f_{\vec{v}}(\vec{v},r) \,,
  \een
\end{subequations}
where $f_{\vec{v}}(\vec{v},r)$ is the velocity distribution in the Galactic frame, generically
defined in the context of the Eddington inversion by \citeeq{eq:v_df}, and $\omega(r)$
ensures the normalization of the distribution to unity over the full available range in
velocity [1 by construction in the Eddington formalism, except if some terms are
  neglected---see discussion below \citeeq{eq:rho_reconstruction}].

Direct searches for WIMP dark matter are typically sensitive to the inverse moment
of the velocity, expressed as the following integral:
\ben
\eta(v_{\rm min}) = \int_{v_{\rm min} \leqslant v \leqslant v_\oplus+v_{\rm esc}}
\mathrm{d}^{3}\vec{v}\, \frac{f_{\vec{v},\oplus}(\vec{v})}{v}\,,
\een
where $f_{\vec{v},\oplus}$ is the WIMP velocity distribution in the rest frame of the Earth. The speed
$v_{\rm min}$ is the minimal speed a DM particle must have to induce a detectable recoil in the
detector. Consequently, low-threshold experiments are sensitive to the high-velocity tail of the
distribution.
For low-mass DM candidates (noted $\chi$ for convenience), with masses much
lower than the target nucleus mass, the minimal speed is $v_{\rm min}\propto 1/m_\chi$ and
can be close to the maximal speed in the laboratory frame $v_{\rm max}=v_\oplus+v_{\rm esc}$,
where the Earth speed in the Galactic frame $v_\oplus$ is close to the Sun speed
$v_\odot\sim 240$~km/s. Giving an accurate description of the tail of the speed distribution
in the Galactic frame is therefore critical, and the regularization of the divergence
associated with $R_{\rm max}$ is crucial in this context. We compare the prediction of the
self-consistent Eddington inversion to the MB approximation. In the context of direct searches,
the MB distribution in the Galactic frame is usually truncated at the escape speed
\cite{LewinEtAl1996}, either
sharply,
\ben
f_{\vec{v}}^{\rm shm}(\vec{v}) = \frac{1}{N_{\rm shm}}\,e^{-v^{2}/v_{\rm circ}^{2}}\,\Theta(v_{\rm esc}-v)\,,
\een
where $\Theta$ is the Heaviside step function, or smoothly,
\ben
f_{\vec{v}}^{\widetilde{\rm shm}}(\vec{v}) = \frac{1}{N_{\widetilde{\rm shm}}}\,
\left(e^{-v^{2}/v_{\rm circ}^{2}}-e^{-v_{\rm esc}^{2}/v_{\rm circ}^{2}}\right)\,.
\een
The respective normalizations are
$N_{\rm shm}=(\pi v_{\rm circ}^{2})^{3/2} [{\rm erf}(z)-2z/\sqrt{\pi}\,\exp(-z^{2})]$ and
$N_{\widetilde{\rm shm}}=(\pi v_{\rm circ}^{2})^{3/2} [{\rm erf}(z)-2z/\sqrt{\pi}\,(1+2z^{2}/3)\,\exp(-z^{2})]$, and $z=v_{\rm esc}/v_{\rm circ}$. Note that the sharply-cut MB distribution is
obviously non-physical due to the step at $v_{\rm esc}$. We consider it nonetheless since it has
been used extensively in the direct searches literature. These deformed MB velocity distributions
are usually dubbed {\em standard halo model} (SHM). In the following, we will pick the values
of $v_{\rm circ}$ at $r=R_\odot$ consistently with the McM17 models used in this study.

Our comparison of $\eta$ for the various cases should not depend significantly on the frame of
reference up to a Galilean shift in velocity, so for simplicity we consider the Galactic frame
rather than the Earth frame (which is the frame relevant for direct searches)---our
$v_{\rm min}$ should thereby be shifted by the Sun speed in the Galactic frame $\sim v_\odot$ to
get values more relevant to direct WIMP searches. We consider the McM17 NFW model for
illustration (see \citesec{app:mass_models}) and the different regularization methods discussed
in \citesec{ssec:rmax}, and assume
\ben
\label{eq:approx_eta}
\eta(v_{\rm min})\simeq \Xi_{-1}(v_{\rm min},v_{\rm esc},R_\odot)\,.
\een
  
We compare the predictions inferred from the SHM and the Eddington inversion for
$\eta(v_{\rm min})$ in the left and right panels of \citefig{fig:one_over_v_vmin} for the isotropic
and Osipkov-Merritt cases, respectively. Generically, predictions derived from the Eddington
inversion differ significantly from that of the SHM over the whole range of $v_{\rm min}$,
as already noticed in the literature \cite{UllioEtAl2001,VergadosEtAl2003,BozorgniaEtAl2013,FornasaEtAl2014,LavalleEtAl2015}---the main difference with previous studies comes from our
rigorous treatment of the issues emphasized in \citesec{sec:issues}, and the selection of stable
configurations only. Differences are especially striking when $v_{\rm min}$ is large due to the
different shapes predicted in the tail of the speed distribution. The smoothly-cut MB distribution
is closer to the Eddington prediction than the sharply-cut MB distribution, but it is also very
discrepant near $v_{\rm esc}$. We also make the comparison with the Osipkov-Merritt models.
\footnote{We do not show the constant-$\beta$ case as the regularization is very similar to the
  isotropic case.} The difference between the SHM and these models are much larger than in the
isotropic case. This is an illustration of the difficulty to regularize the Osipkov-Merritt
models, for which none of the prescriptions are fully satisfactory (see \citesec{ssec:rmax}).
Either the divergence is not removed ($R_{\rm max}\rightarrow\infty$ case) or the underlying
density profile is significantly modified.

Thus, irrespective of the regularization and the anisotropy, the prediction of the self-consistent
approach systematically differs from the SHM. We are able to quantify the theoretical
uncertainties associated with the treatment of the divergence, which is especially important for
large values of $v_{\rm min}$. This is critical for low-mass DM candidates in direct searches.
\begin{figure*}[!t]
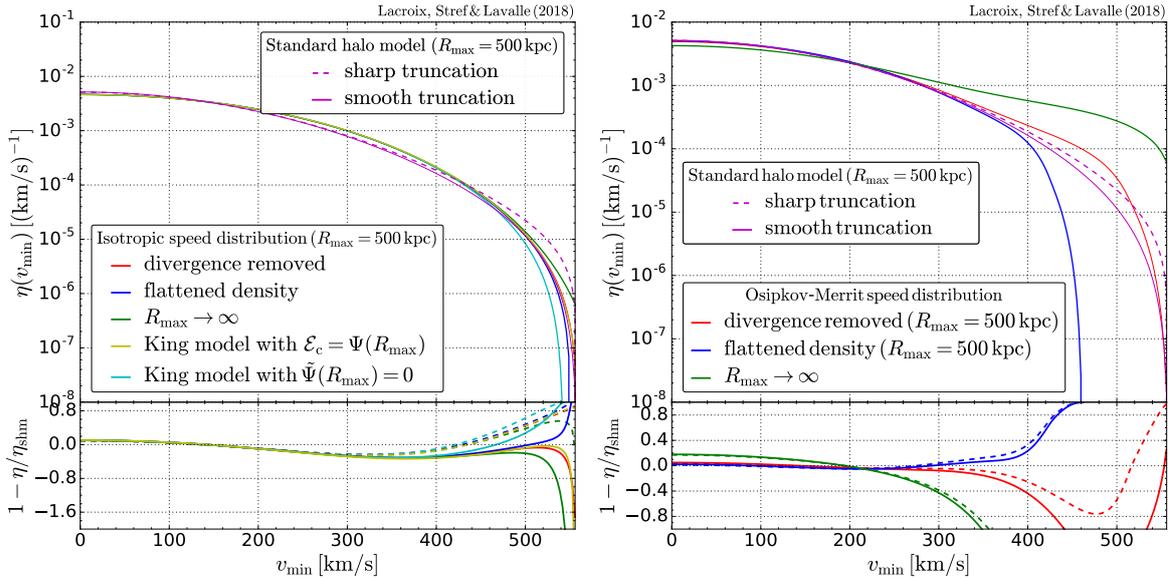

  \centering
  \includegraphics[width=0.495\textwidth]{{{fig_eta_isotropic_king_models}}}
  \includegraphics[width=0.495\textwidth]{{{fig_eta_vmin_om}}}
  \caption{\small \textbf{Left panel:} $\eta$ integral as a function of $v_{\mathrm{min}}$. The
    various curves shown are the sharply-cut SHM (solid magenta), the smoothly-cut SHM (dashed
    magenta), and the predictions of the Eddington formalism for an isotropic system, with the
    regularizations of the phase-space divergence discussed in \citesec{ssec:rmax}, namely setting
    $R_{\mathrm{max}}$ to infinity (green), removing the diverging term (red), modifying the
    density profile (blue), regularizing \`a la King with ${\cal E}_c=\Psi(R_{\rm max})$ (yellow)
    or with $\tilde\Psi(R_{\rm max})=0$ (cyan). \textbf{Right panel:} Same as left panel, for the
    Osipkov-Merritt model.}
  \label{fig:one_over_v_vmin}
\end{figure*}
For the sake of completeness, we also compare the Eddington inversion and MB results obtained
for the observables proportional to $\bar \eta \equiv \Xi_{-1}(0,v_{\rm max},r)$, which
could be related to the capture of DM in stars or planets (\eg\ \cite{PressEtAl1985,Gould1987,SalatiEtAl1989,BouquetEtAl1989,BouquetEtAl1989a,Kouvaris2008,BertoneEtAl2008}), and to
$\langle v \rangle = \Xi_{1}(0,v_{\rm esc},r)$, which could be related to the microlensing event
rate of compact DM objects (\eg\ \cite{Griest1991,Green2017a})---the latter is simply the mean
speed across the Galaxy.

\begin{figure*}[!t]
  \centering
  \includegraphics[width=0.495\textwidth]{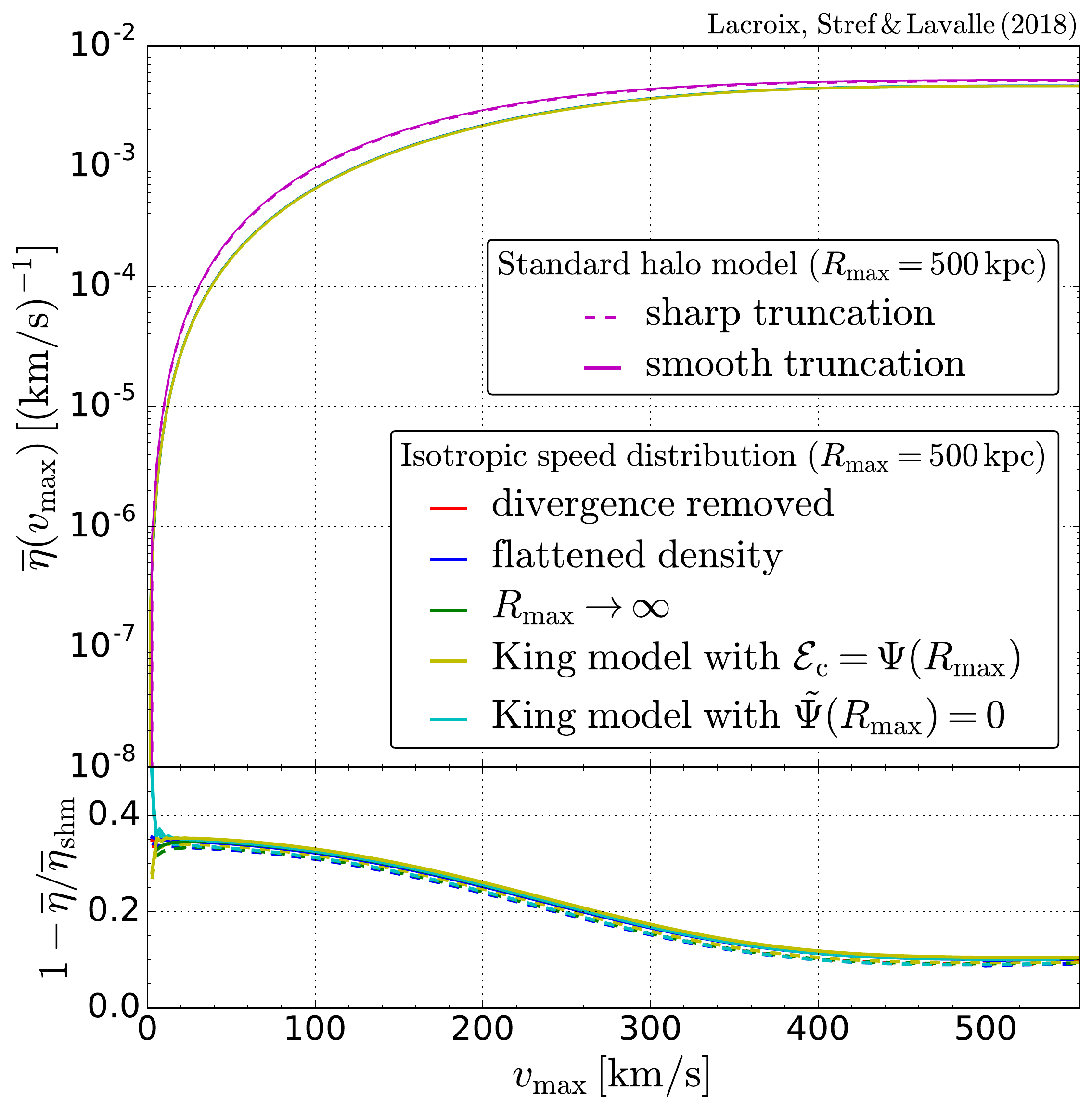}
  \includegraphics[width=0.495\textwidth]{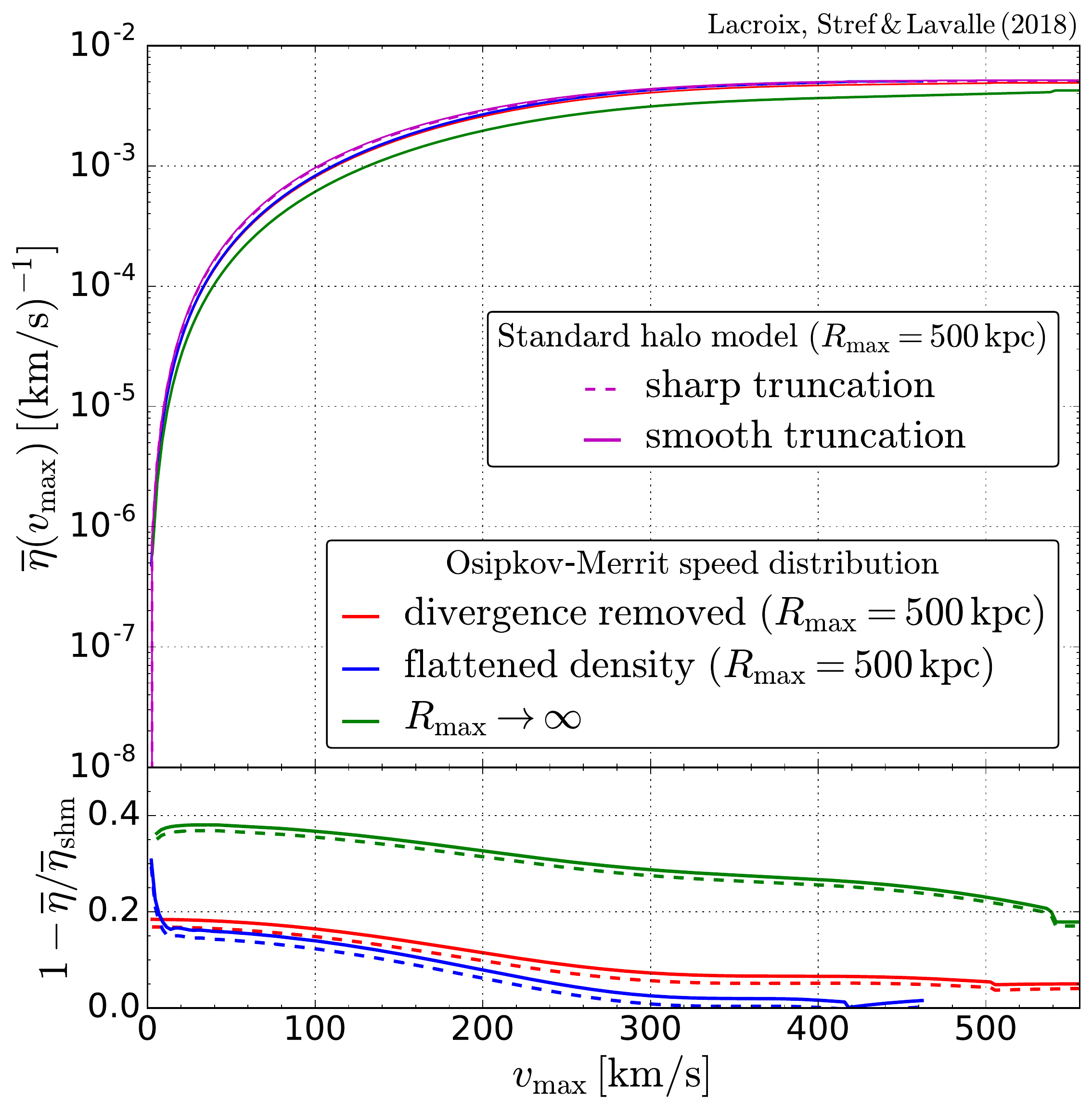}
  \caption{\small Same as Fig.~\ref{fig:one_over_v_vmin} for $\bar \eta \equiv
    \Xi_{-1}(0,v_{\rm max},R_\odot)$.}
  \label{fig:other_1_to_vmoments}
\end{figure*}

\begin{figure*}[!t]
  \centering
  \includegraphics[width=0.495\textwidth]{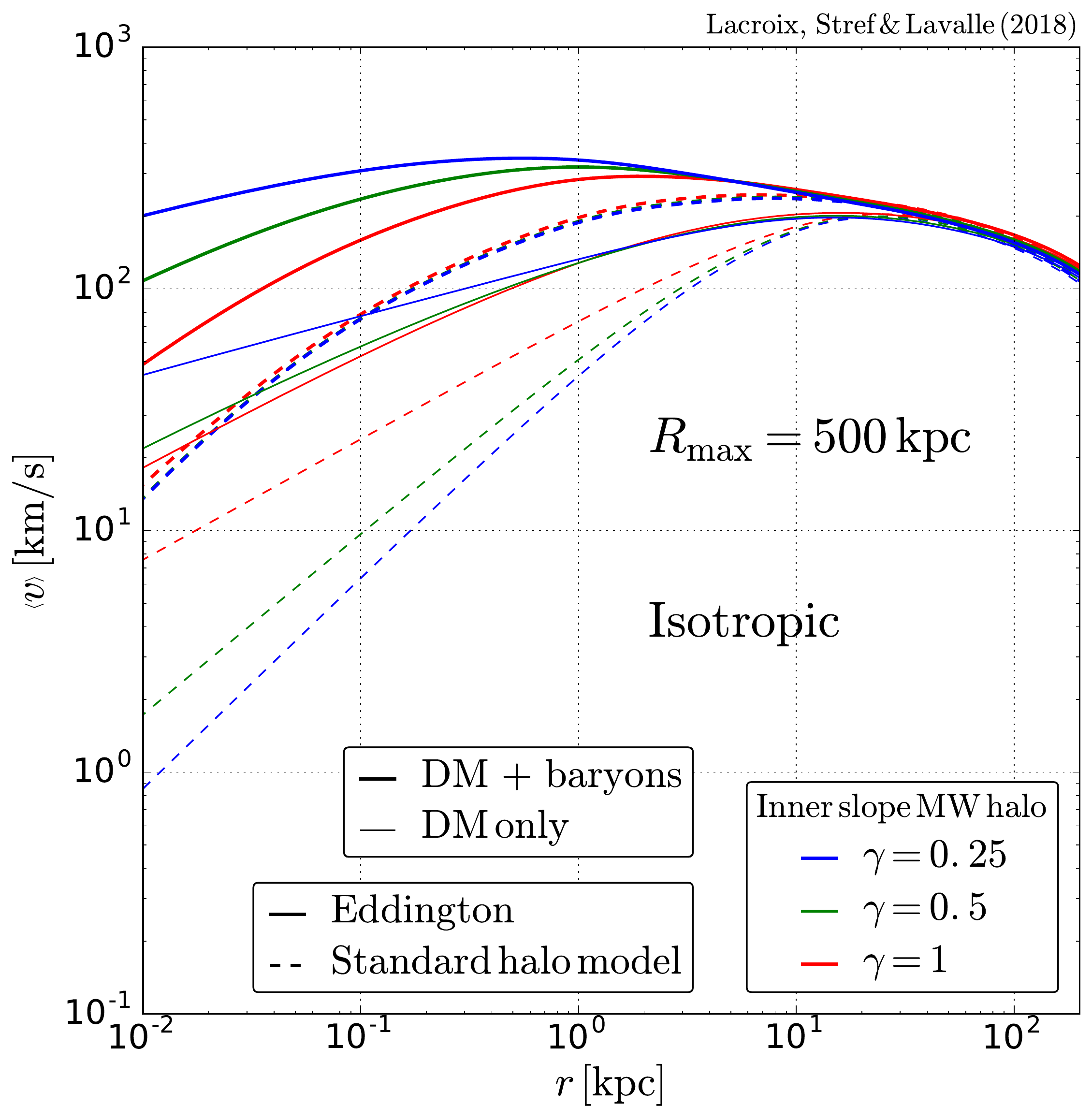}
  \includegraphics[width=0.495\textwidth]{{{fig_vmean_om_new}}}
  \includegraphics[width=0.495\textwidth]{{{fig_vmean_beta-0.3_new}}}
  \caption{\small Mean speed profiles for the Standard Halo Model and the Eddington formalism for
    the isotropic (top left panel), Osipkov-Merritt (top right panel) and $\beta_{0} = -0.3$
    (bottom panel) cases. Here we show the DM-only (thin line) and DM+baryons (thick line) cases,
    for the McM17 mass models providing well-behaved Eddington-inverted DFs.}
  \label{fig:other_vmoments}
\end{figure*}

Our results are illustrated in Figs.~\ref{fig:other_1_to_vmoments} and
\ref{fig:other_vmoments} for $\bar{\eta}$ (for which we set $r=R_\odot$) and $\langle v \rangle$,
respectively. For $\bar \eta$, we see significant differences between the Eddington inversion
and the Maxwellian approximation, decreasing from $\sim 40\%$ to $\sim 10\%$ as $v_{\rm max}$ spans
the full dynamical range---we also see that isotropic DFs are poorly sensitive to the radial
cutoff treatment, in contrast to anisotropic DFs, where radial orbits come into play. 
\newchange{
For the mean speed $\langle v \rangle$, the only regions where the Maxwellian approximation provides results similar to the Eddington inversion are 
the outer parts of the Galaxy. The departure between the two prediction increases as the radius gets smaller, with up to an order of magnitude of difference at the center of the Galaxy. This should therefore be considered seriously in
predictions of related observables.}
The negative $\beta$ case leads to a mean speed
curve closer to the Maxwellian case, as expected for more circular orbits (the mean speed then tends
to the circular speed). Note that the Maxwellian results are obviously similar for all Galactic
models when both the DM and baryons are included, as these models are constrained from rotation
curves; they consequently separate from each other when only the DM halo is considered (the DM mass
profiles may vary significantly in regions dominated by the baryons). The results
obtained for the moments of the relative speed in \citesec{ssec:idlike} exhibit the same behavior.

\subsection{Indirect-search-like observables}
\label{ssec:idlike}
Other DM-related signals are related to moments (or inverse moments) of the relative
speed instead of the speed. This concerns signals related to two-body processes, whose
most striking example is the self-annihilation of DM. We therefore define a new moment
function for the relative speed $\vec{v}_{\rm r}=\vec{v}_2-\vec{v}_1$ between two DM particles,
\begin{subequations}
  \label{eq:vr_moments}
  \ben
  \Pi_{n}(v_{\rm min},v_{\rm max},r)&\equiv& \kappa^{-1}(r)
  \int_{v_{\rm min}}^{v_{\rm max}} {\rm d}^{3}\vec{v}_{1}
  \int_{v_{\rm min}}^{v_{\rm max}} {\rm d}^{3}\vec{v}_{2}\,
  |\vec{v}_{\rm r}|^{n}\,
  f_{\vec{v}}(\vec{v}_{1},r)\,
  f_{\vec{v}}(\vec{v}_{2},r)\\
  \kappa(r)&\equiv&
  \int {\rm d}^{3}\vec{v}_{1} \int {\rm d}^{3}\vec{v}_{2}\,
  f_{\vec{v}}(\vec{v}_{1},r) \, f_{\vec{v}}(\vec{v}_{2},r)\,,
  \een
\end{subequations}
where the velocity distribution $f_{\vec{v}}$ is conventionally defined by \citeeq{eq:v_df}
in the context of Eddington's inversion formalism, and the function $\kappa(r)$ ensures
the correct normalization to unity in the relevant range of individual speed
[1 by construction in the Eddington formalism, except if some terms are
  neglected---see discussion below \citeeq{eq:rho_reconstruction}].

Indirect searches for self-annihilating DM are sensitive to the following moments
\ben
\label{eq:moments_vr}
\langle |\vec{v}_{\rm r}|^{n} \rangle (r) = \Pi_{n}(0,v_{\rm esc},r)
\een
Searches for $p$-wave annihilation typically probe the (relative) velocity dispersion ($n=2$),
though in some interaction models the annihilation cross-section can be modified by
non-perturbative effects \cite{HisanoEtAl2005} that lead to the so-called Sommerfeld enhancement,
which induces a dependency on the $n=-1$ moment---as well as the $n=-2$ moment at resonances. Note
that in practice it proves convenient to perform the following change of variable to express the
integrals in terms of the center-of-mass velocity $\vec{v}_{\rm c}$ and relative velocity
$\vec{v}_{\rm r}$ (\eg\ \cite{GondoloEtAl1991}):
\ben
\left\{
\begin{array}{ll}
  \vec{v}_{\rm c} &= (\vec{v}_{1}+\vec{v}_{2})/2\\
  \vec{v}_{\rm r} &= \vec{v}_{2}-\vec{v}_{1}.
  \end{array}
\right.
\een
As a result, \citeeq{eq:moments_vr} can be rewritten
\ben
\langle v_{\rm r}^{n} \rangle =
\int {\rm d}^{3}\vec{v}_{\rm r}\,|\vec{v}_{\rm r}|^{n}\,F_{\rm r}(\vec{v}_{\rm r},r)\,,
\een
where $F_{\rm r}$ is the relative velocity DF, which is defined as
\ben
\label{eq:def_vr_df}
F_{\rm r}(\vec{v}_{\rm r},r) \equiv \kappa^{-1}(r)
\int {\rm d}^{3}\vec{v}_{\rm c}\,f_{\vec{v}}(\vec{v}_{1},r)\,f_{\vec{v}}(\vec{v}_{2},r)\,.
\een
The full derivation of $F_{\rm r}(\vec{v}_{\rm r},r)$ is given in \citeapp{app:relative_dist_DF} in
the Eddington formalism and its anisotropic extensions discussed above. To our knowledge, the
computation of $F_{\rm r}(\vec{v}_{\rm r},r)$ in the general anisotropic case is an original result.
An alternative treatment for the Osipkov-Merritt models is presented in Ref.~\cite{PetacEtAl2018}.

%


We show the predictions for the relative speed moments inferred from the 
\newchange{(smoothly-truncated)} 
SHM and the isotropic
Eddington inversion in \citefig{fig:moments_vr}. Following our discussion regarding the stability
of the DFs, we only consider here mass models leading to stable solutions of the Boltzmann
equation. The velocity distribution in the Eddington case have been computed \textit{without} the
diverging term, \ie~using \citeeq{eq:df_without_divergence}. We recall that this is in practice
similar to assuming a flattened density profile at the outskirts of the halo, as in
\citeeq{eq:alt_density}. One can see in \citefig{fig:moments_vr} that, for both the SHM and the
Eddington model, the moments with and without baryons converge at large radii. This is because the
total mass, and therefore the gravitational dynamics, is then fully dominated by DM, and
baryons become irrelevant. Though similar in shape, predictions from the two models are numerically
quite different. For the $n>0$ moments, the Eddington model's predictions typically exceed the
SHM's. At the center of the Galaxy, the two models differ by at least an order of magnitude, up to
three orders of magnitude. The hierarchy of the moments with respect to the value of DM inner
slope $\gamma$ is also reverted. While the cuspiest mass model ($\gamma=1$) leads to the largest
prediction for the SHM, it is the model closest to the core ($\gamma=0.25$) that dominates the
Eddington result. 
We stress that even locally at $r=R_\odot\sim 8\,\rm kpc$, and for all $n$, there are
sizable differences between the Eddington formalism and the Maxwell-Boltzmann approximation.
Therefore, since the Eddington formalism turns out to better capture the dynamical properties of
the DM halo than the SHM \cite{LacroixEtAl2018}, the latter should only be used to make very rough
estimates of $p$-wave annihilating DM signals, even when isotropy is assumed.

We also compared the (isotropic) SHM with some of the anisotropic extensions of the Eddington
formalism. The prediction of the Osipkov-Merritt model is shown in \citefig{fig:moments_vr_om} for
a particular choice of the anisotropy radius $r_{\rm a}=r_{\rm s}$. Note that the value of $r_{\rm s}$
depends on the underlying mass model (see \citetab{tab:dm_mass_models}). The result is close to the
isotropic case at radii $r\ll r_{\rm a}$, as expected from the behavior of the anisotropy parameter
\citeeq{eq:beta_om}. At large radii however, the slope of the moments steepens significantly. The
steepening starts roughly where $r\simeq r_{\rm a}$ which is where the system begins to be strongly
anisotropic. We stress again the fact the regularization of the diverging term changes considerably
the underlying density profile in the Osipkov-Merritt case, as seen from
\citefig{fig:reconstructed_rho}. The behavior of $\langle v_{\rm r}^{n}\rangle $ beyond
$r=r_{\rm a}$ should therefore be treated with caution. We also studied the constant anisotropy
case, focusing on $\beta_{0}=-0.3$. We considered a negative anisotropy to get a well-defined DF
for all the mass models of relevance here. The corresponding relative speed moments are shown in
\citefig{fig:moments_vr_beta}. They differ from the isotropic ones at all radii, unlike the
Osipkov-Merritt ones, which is not surprising since the constant anisotropy is non zero everywhere.

Regardless of the assumption made on the anisotropy, the Eddington formalism generically predicts
huge differences with respect to the SHM. The various anisotropic models we used allow us to
bracket the theoretical uncertainty on the Eddington method. 

\begin{figure*}[!t]
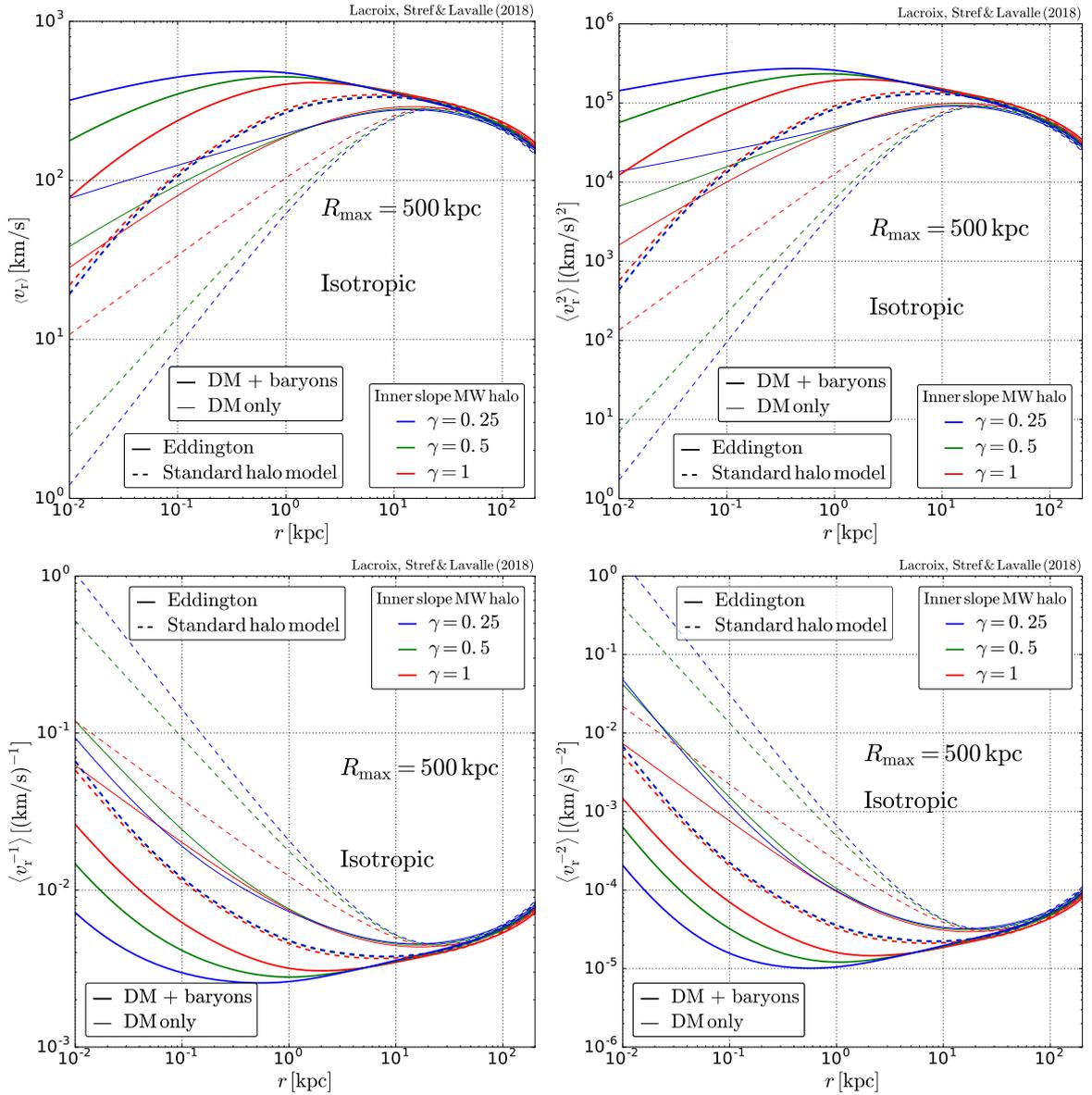

  \centering
  \includegraphics[width=0.495\textwidth]{{{moment_vr1_isotropic_v2}}}
  \includegraphics[width=0.495\textwidth]{{{moment_vr2_isotropic_v2}}}
  \includegraphics[width=0.495\textwidth]{{{moment_vr-1_isotropic_v2}}}
  \includegraphics[width=0.495\textwidth]{{{moment_vr-2_isotropic_v2}}}
  \caption{\small Moments of the relative velocity distribution, for the Standard Halo Model and
    the Eddington formalism (isotropic case). Here we show the DM-only (thin line) and
    DM+baryons (thick line) cases, for several mass models from Ref.~\cite{McMillan2017}.}
  \label{fig:moments_vr}
\end{figure*}

\begin{figure*}[!t]
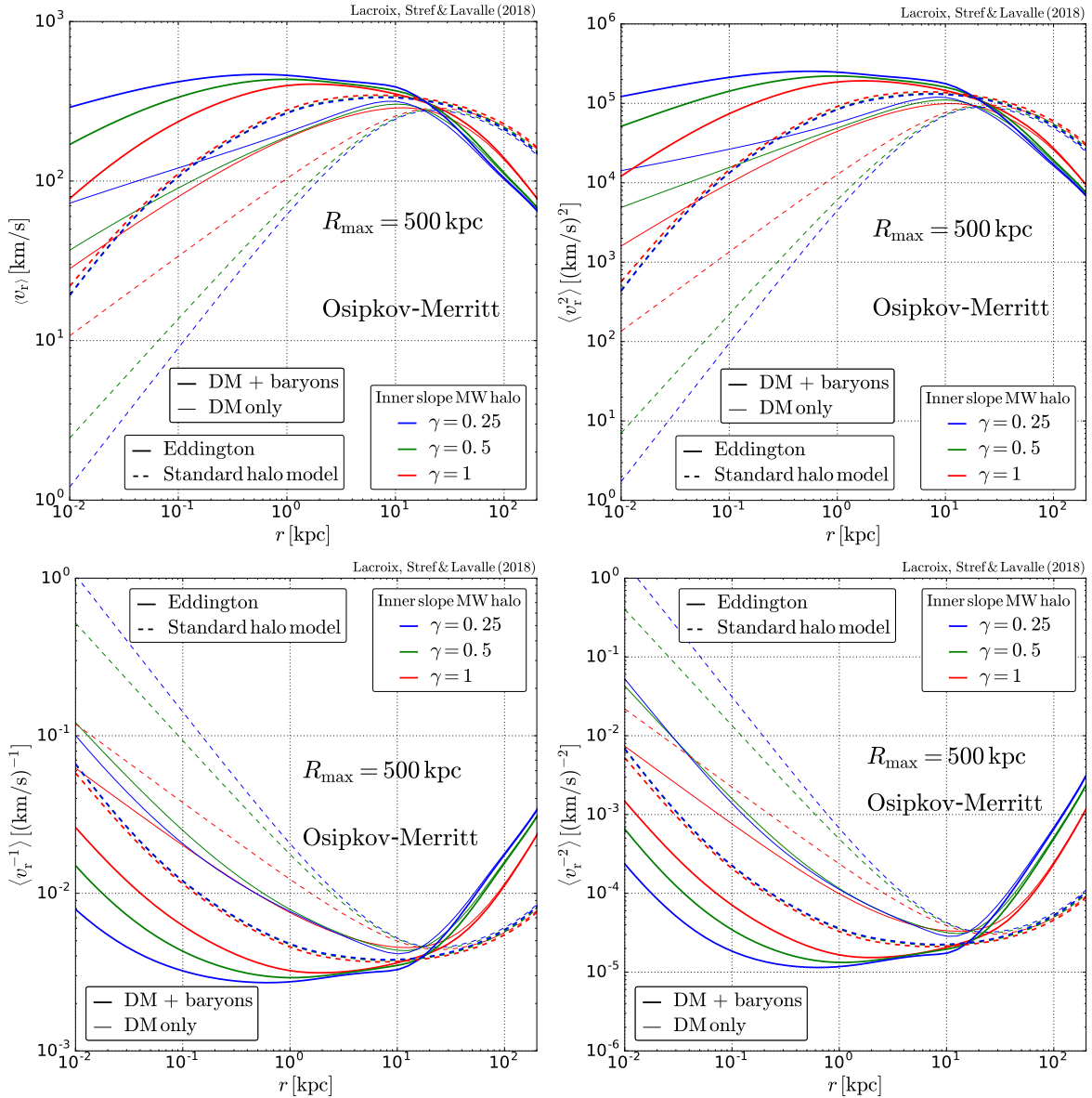

  \centering
  \includegraphics[width=0.495\textwidth]{{{moment_vr1_om_v2}}}
  \includegraphics[width=0.495\textwidth]{{{moment_vr2_om_v2}}}
  \includegraphics[width=0.495\textwidth]{{{moment_vr-1_om_v2}}}
  \includegraphics[width=0.495\textwidth]{{{moment_vr-2_om_v2}}}
  \caption{\small Same as \citefig{fig:moments_vr}, for the Osipkov-Merritt model with
    $r_{\rm a}=r_{\rm s}$.}
  \label{fig:moments_vr_om}
\end{figure*}

\begin{figure*}[!t]
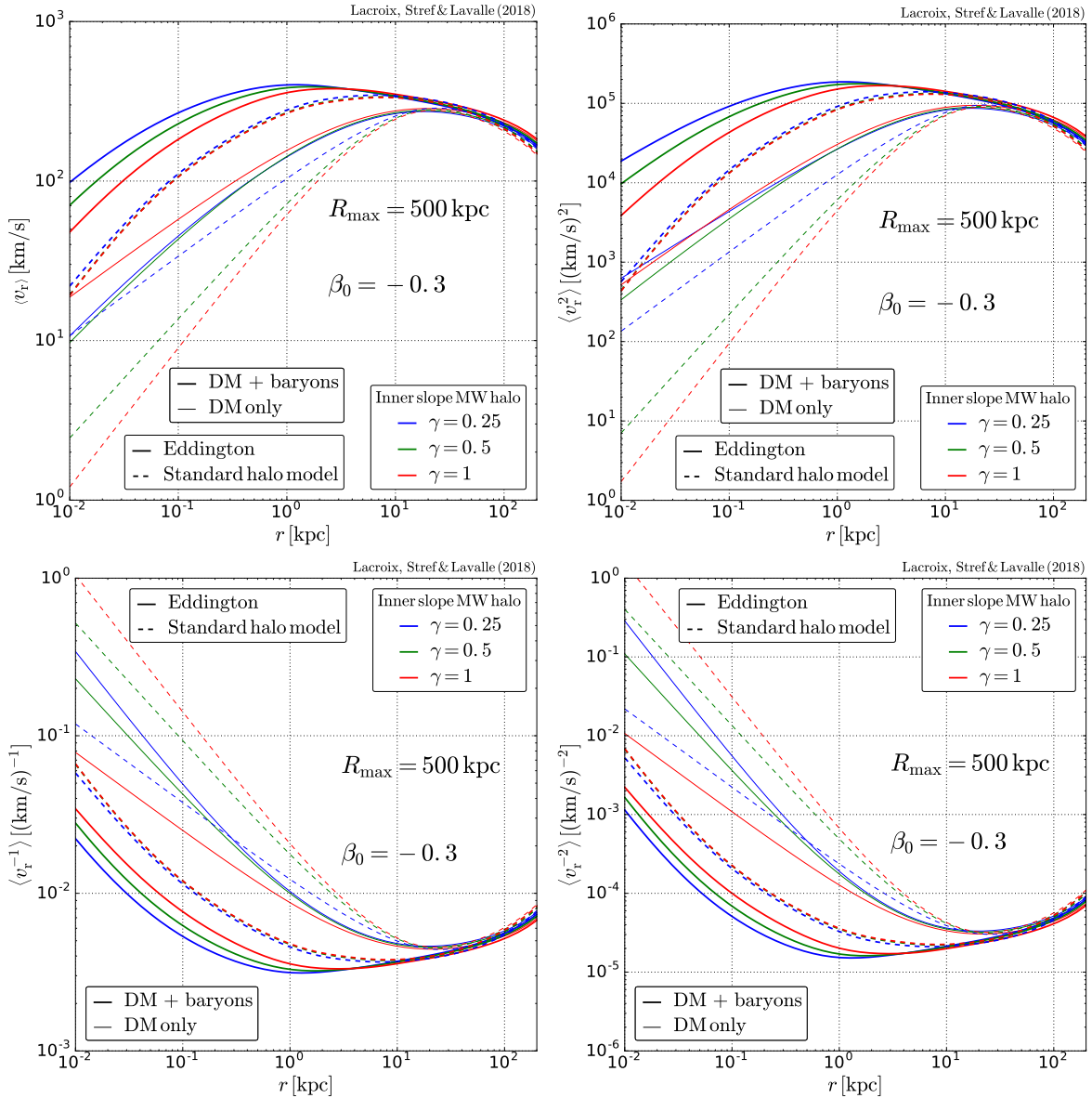

  \centering
  \includegraphics[width=0.495\textwidth]{{{moment_vr1_cst_beta_v2}}}
  \includegraphics[width=0.495\textwidth]{{{moment_vr2_cst_beta_v2}}}
  \includegraphics[width=0.495\textwidth]{{{moment_vr-1_cst_beta_v2}}}
  \includegraphics[width=0.495\textwidth]{{{moment_vr-2_cst_beta_v2}}}
  \caption{\small Same as \citefig{fig:moments_vr}, for the constant-$\beta$ model with
    $\beta_{0}=-0.3$.}
  \label{fig:moments_vr_beta}
\end{figure*}
  
\section{Conclusion}
\label{sec:concl}
In this paper, we have reviewed the Eddington inversion formalism, and a few of its generic
extensions to anisotropic systems. This formalism is powerful to consistently include the dynamical
correlations featured by a self-gravitating system in the DM-search-related velocity-dependent
observables, from a mass model constrained on real data. It represents a strong improvement over
the Maxwellian approximation from both the theoretical and quantitative points of view, and should
therefore become a ``next-to-minimal'' standard approach to refine the predictions and better
quantify the dynamical uncertainties in DM-search predictions pertaining to (sub)Galactic scales.
It is also more appealing theoretically than blindly using ad-hoc fits from cosmological
simulations, which likely hide environmental dependencies or other artifacts.
Though not as evolved nor as adaptable to a large variety of potential-density pairs as the
action-angle formalism \cite{SandersEtAl2016},
Eddington's inversion method still provides a decent description of galactic DM halos
\cite{LacroixEtAl2018} from a minimal set of physical assumptions and a moderate level
of technicalities\change{--pending the breakdown of some assumptions (\eg~spherical symmetry,
  steady state, smoothness of the dark halo, \etc.) that induces additional systematic errors which
  remain to be quantified}. Inspecting the self-consistency of this approach is therefore
particularly important at the time of a boost in astrometric precision made
possible by the Gaia mission \cite{BrownEtAl2018}, which should provide much better constraints on
the DM content of the MW and its satellites. This is of special relevance in the context of
intense DM searches, as the Eddington-like inversion methods are well suited to better control and
further reduce the astrophysical uncertainties in the signal predictions---\eg\ in direct
\cite{UllioEtAl2001,CatenaEtAl2012,FornasaEtAl2014}, indirect \cite{FerrerEtAl2013,Hunter2014,
  BoddyEtAl2017,PetacEtAl2018}, or combined \cite{CerdenoEtAl2016} WIMP searches, but not only.

After carefully inspecting the Eddington inversion formalism in \citesec{sec:edd}, however, we
noticed several theoretical issues related to (i) the radial boundary of the dark halo,
important to make sense of the constraints on the escape speed \cite{LavalleEtAl2015}, and (ii)
to the stability of the phase-space DF, which have been overlooked in the DM-related literature,
but which are actually expected to arise very often when Eddington inverting Galactic mass models
with a baryonic component. We have described and addressed these issues in \citesec{sec:issues},
and provided generic methods to cure some of the potential inconsistencies. For the divergence
induced in the phase-space DF in the limit of $v\to v_{\rm esc}$ (see \citesec{ssec:rmax}), after
explaining why the diverging term $\propto 1/\sqrt{\cal E}$ should actually not be dropped, we
defined two ways of getting a non-anomalous phase-space DF without spoiling too much the initial
mass model, based either on {\em a priori} modifications of the DM profile or on new converging
ans\"atze for the DF itself. Properly describing the boundary of the system is particularly
important to characterize
the theoretical uncertainties affecting DM-search observables depending on the high-velocity tail
of the DM velocity distribution, like the direct detection rate of GeV- or subGeV-mass WIMPs.
The proposed regularization methods proved efficient in all cases, except in the case of the
anisotropic Osipkov-Merritt model, which cannot consistently accommodate radial boundaries. As for
the stability issue (see \citesec{ssec:gamma}), we recovered stability criteria for
both the isotropic case [see \citeeq{eq:stability2}] and the anisotropic case [see
  \citeeq{eq:stability_anisotropy}, a criterion for stability against radial perturbations only].
The former criterion could be complemented by \citeeq{eq:stability3} to ensure a smoother
velocity distribution, while this restricts the phase-space volume beyond the requirement of
stability only. We also analyzed these stability conditions to get selection criteria for a Galactic
halo based on its relative baryonic content---see \citeeq{eq:stability_baryons} and
\citefig{fig:contours}. This allows one to quickly check whether \change{one's} favorite Galactic
mass model is Eddington invertible or not. The main conclusion of this part is that Eddington's
inversion (and its anisotropic extensions) cannot blindly apply to any density-potential pair.
Particular attention should be given to moderately cuspy or cored halo profiles, which are more
likely to exhibit ill-defined DFs.

Finally, we have explicitly computed some DM-search velocity-dependent observables to
explicitly compare the Eddington inversion predictions with those derived in the Maxwellian
approximation, in the framework of the McM17 constrained Galactic mass models \cite{McMillan2017}.
In particular, we have computed observables that depend on the speed moments
(and inverse moments), and on the relative speed moments (and inverse moments). The former ones
regard the direct WIMP detection rate, and the latter ones regard $p$-wave self-annihilation of
DM. For the self-annihilation case, we have derived a convenient way to express the relative
velocity distribution function for anisotropic systems, reviewed in
\citeapp{app:relative_dist_DF}. We have seen that the differences are quite sizable in all
observables, which somewhat quantifies the associated level of theoretical uncertainties.
We will actually show in a forthcoming study that the Eddington inversion methods provide a
significantly better description of the true DF than the Maxwellian approximation in
zoomed-in cosmological simulations with baryons \cite{LacroixEtAl2018}. This further motivates
applications of this approach to exploit the upcoming Gaia-constrained mass models in the
context of DM searches.

\acknowledgments{
  The PhD grant of MS is funded by the OCEVU Labex (ANR-11-LABX-0060), which also provided
  financial support to this project. We also benefited from financial support from the theory
  project {\em Galactic Dark Matter} funded by CNRS-IN2P3. We further acknowledge support from the
  European Union's Horizon 2020 research and innovation program under the Marie
  Sk\l{}odowska-Curie grant agreements No 690575 and  No 674896; beside recurrent institutional
  funding by CNRS-IN2P3 and the University of Montpellier.
}

\appendix
\section{Galactic mass models used in this study}
\label{app:mass_models}
For definiteness, we used a selection of mass models from Ref.~\cite{McMillan2017}
(the McM17 model henceforth), featuring a
stellar bulge, two stellar disks, two gas disks and a DM halo. The bulge profile reads
\ben
\rho_\mathrm{b}=\frac{\rho_{0,\mathrm{b}}}{(1+r'/r_0)^\alpha}\;
\textrm{exp}\left[-\left(r'/r_{\rm b}\right)^2\right],
\label{eq:bulge}        
\een
where
\ben
r' = \sqrt{R^2 + (z/q)^2}.
\een
The variable $q$ determines the oblateness of the bulge, $\rho_{0,\mathrm{b}}$ is a scale density,
and $r_{0}$ and $r_{\rm b}$ are scale lengths. The stellar disks are modeled by exponential
profiles:
\ben
\label{eq:disk}
\rho_{*,\mathrm{d}}(R,z)=\frac{\Sigma_{0}}{2z_\mathrm{d}}\;\textrm{exp}
\left(-\frac{|z|}{z_\mathrm{d}}-\frac{R}{R_\mathrm{d}}\right),
\een
with scale height $z_\mathrm{d}$, scale length $R_\mathrm{d}$ and central surface density
$\Sigma_{0}$. The HI and H$_{2}$ gas disks are described by
\ben
\label{eq:gasdisk}
\rho_\mathrm{g,d}(R,z)=\frac{\Sigma_{0}}{4z_\mathrm{d}}\;
\exp\left(-\frac{R_{\rm m}}{R}-\frac{R}{R_\mathrm{d}}\right)\; {\rm sech}^2(z/2z_{\mathrm{d}}).
\een
Finally, the DM halo is characterized by a generalized $\alpha\beta\gamma$ profile \cite{Zhao1996}
\ben
 \rho_{\mathrm{DM}}(x) = \rho_{\rm s}\,x^{-\gamma}\left(1+x^{\alpha}\right)^{(\gamma-\beta)/\alpha}
 \label{eq:halo}
\een    
with $x=r/r_{\mathrm{s}}$, where $r_{\rm s}$ is the scale radius. An NFW profile is recovered
with $(\alpha,\beta,\gamma)=(1,3,1)$. The author of Ref.~\cite{McMillan2017} uses instead
$\alpha=1$ and $\beta=3$, and fits the data for specific values of $\gamma$. For completeness, we
also give the NFW gravitational potential:
\ben
\phi_{\rm nfw}(r) = -4\,\pi\,G\,\rho_{\rm s}\,r_{\rm s}^2\,x^{-1}\,\ln\left( 1+x\right)\,.
\een

The best-fit parameters for the McM17 model are given in Tables \ref{tab:dm_mass_models}
and \ref{mass_model_table}.

\begin{table}
\caption{\label{tab:dm_mass_models}\small Dark matter mass models from Ref.~\cite{McMillan2017}.}
\centering
\renewcommand{\arraystretch}{1.3}%
\begin{tabular}{c c c c c}
$(\alpha,\beta,\gamma)$ & $\rho_{\rm s}$ & $r_{\rm s}$ & $\rho_{\odot}$ & $R_{\odot}$\\
 & $\rm [M_{\odot}/kpc^{3}]$ & $\rm [kpc]$ & $\rm [M_{\odot}/kpc^{3}]$ & $\rm [kpc]$\\
 \hline
(1,\,3,\,0) & $9.09\times10^{7}$ & 7.7 & $1.03\times10^{7}$ & 8.21 \\
\hline
(1,\,3,\,0.25) & $5.26\times10^{7}$ & 9.6 & $1.01\times10^{7}$ & 8.20 \\
\hline
(1,\,3,\,0.5) & $3.19\times10^{7}$ & 11.7 & $1.01\times10^{7}$ & 8.21 \\
\hline
(1,\,3,\,1) & $8.52\times10^{6}$ & 19.6 & $1.01\times10^{7}$ & 8.21 \\
\end{tabular}
\end{table}

\begin{table}[h!]
  \caption{\label{mass_model_table}\small Baryonic mass model from Ref.~\cite{McMillan2017}, for
    the NFW DM profile ($\gamma=1$, see \citetab{tab:dm_mass_models}).}
{\renewcommand{\arraystretch}{1.3}%
\begin{tabular}{ccc}
 & Parameter & Value\\
 \hline
Bulge & $\rho_{0,\mathrm{b}}$ & $9.84 \times 10^{10}\, \mathrm{M_{\odot}\, kpc^{-3}}$\\
& $r_{0,\mathrm{b}}$ & $0.075\, \mathrm{kpc}$\\
& $r_{\rm b}$ & $2.1\, \mathrm{kpc}$\\
& $q$ & $0.5$\\
& $\alpha$ & $1.8$\\
\hline
Stellar disks & $\Sigma_{0,\mathrm{thin}}$ & $8.96 \times 10^{8}\, \mathrm{M_{\odot} \, kpc^{-2}}$\\
& $R_{\mathrm{d,thin}}$ & $2.5\, \mathrm{kpc}$\\
& $z_{\mathrm{d,thin}}$ & $0.3\, \mathrm{kpc}$\\
& $\Sigma_{0,\mathrm{thin}}$ & $1.83 \times 10^{8}\, \mathrm{M_{\odot} \, kpc^{-2}}$\\
& $R_{\mathrm{d,thick}}$ & $3.02\, \mathrm{kpc}$\\
& $z_{\mathrm{d,thick}}$ & $0.9\, \mathrm{kpc}$\\
\hline
Gas disks& $\Sigma_{0,\mathrm{HI}}$ & $5.31 \times 10^{7}\, \mathrm{M_{\odot} \, kpc^{-2}}$\\
& $R_{\mathrm{d,HI}}$ & $7\, \mathrm{kpc}$\\
& $R_{\mathrm{m,HI}}$ & $4\, \mathrm{kpc}$\\
& $z_{\mathrm{d,HI}}$ & $0.085\, \mathrm{kpc}$\\
& $\Sigma_{0,\mathrm{H_{2}}}$ & $2.18 \times 10^{9}\, \mathrm{M_{\odot} \, kpc^{-2}}$\\
& $R_{\mathrm{d,H_{2}}}$ & $1.5\, \mathrm{kpc}$\\
& $R_{\mathrm{m,H_{2}}}$ & $12.\, \mathrm{kpc}$\\
& $z_{\mathrm{d,H_{2}}}$ & $0.045\, \mathrm{kpc}$\\
\end{tabular}
}
\end{table}

\section{Relative velocity distribution function}
\label{app:relative_dist_DF}
Here we provide the full derivation of the moments of the relative speed relevant for
DM observables involving two-body processes. The $n^{\rm th}$ moment reads
\ben
\langle v_{\rm r}^{n} \rangle =
\int \! {\rm d}^{3}\vec{v}_{\rm r} \, |\vec{v}_{\rm r}|^{n} \, F_{\rm r}(\vec{v}_{\rm r},r)\,,
\een
where the relative velocity distribution function at Galactic radius $r$ already
introduced in \citeeq{eq:def_vr_df} is expressed as an integral over the center-of-mass
velocity as follows:
\ben
F_{\rm r}(\vec{v}_{\rm r},r) = \kappa^{-1}(r) \int \! \mathrm{d}^{3}\vec{v}_{\rm c} \,
f_{\vec{v}}(\vec{v}_{1},r)\,f_{\vec{v}}(\vec{v}_{2},r).
\een
The normalization function $\kappa(r)$ has been defined in \citeeq{eq:vr_moments}.
\subsection{Isotropic system}
For an isotropic system, $f_{\vec{v}}(\vec{v}) = f_{\vec{v}}(v)$. Going from $\vec{v}_{1}$ and
$\vec{v}_{2}$ to center-of-mass and relative velocities,
\ben
\label{eq:vc_vr}
\left\{
\begin{array}{ll}
  \vec{v}_{\rm c} &= (\vec{v}_{1}+\vec{v}_{2})/2\\
  \vec{v}_{\rm r} &= \vec{v}_{2}-\vec{v}_{1},
  \end{array}
\right.
\een
one can write
\ben
\langle v_{\rm r}^{n} \rangle = 8 \, \pi^{2 }
\int_{v_{\rm r}^{\mathrm{min}}}^{v_{\rm r}^{\mathrm{max}}} {\rm d}v_{\rm r} \, v_{\rm r}^{2}
\, F_{\rm r}(\vec{v}_{\rm r},r) \,\vec{v}_{\rm r}^{n}\,.
\een
Let us define the angle
\begin{subequations}
  \label{eq:v_angles}
  \ben
  \theta &\equiv& (\vec{v}_{\rm c},\vec{v}_{\rm r}) \\
  \text{and}\; \mu &\equiv& \cos \theta\,.
  \een
\end{subequations}
From \citeeq{eq:vc_vr}, we get
\ben
\left\{
\begin{array}{ll}
  \vec{v}_{1} &= \vec{v}_{\rm c} - \dfrac{\vec{v_{\rm r}}}{2}\\
  \vec{v}_{2} &= \vec{v}_{\rm c} + \dfrac{\vec{v_{\rm r}}}{2},
  \end{array}
\right.
\een
hence the associated moduli
\begin{subequations}
\label{eq:v1_v2}
\ben
v_{1} &=& |\vec{v}_{1}| = \sqrt{v_{\rm c}^{2} + \dfrac{v_{\rm r}^{2}}{4} - v_{\rm c}\,v_{\rm r}\,\mu}\\
v_{2} &=& |\vec{v}_{2}| = \sqrt{v_{\rm c}^{2} + \dfrac{v_{\rm r}^{2}}{4} + v_{\rm c}\,v_{\rm r}\,\mu}.
\een
\end{subequations}
In the Galactic frame, we have $v_{1} \leqslant v_{\mathrm{esc}}$, which gives the lower and upper
bounds on $\mu$, $| \mu | \leqslant \mu_{0}$, where
\ben
\mu_{0}(r,v_{\rm r},v_{\rm c}) =
\dfrac{v_{\mathrm{esc}}^{2} - v_{\mathrm{c}}^{2} - v_{\mathrm{r}}^{2}/4 }{v_{\mathrm{c}} v_{\mathrm{r}}}.
\een
\citeeq{eq:vc_vr} also implies that $0 \leqslant v_{\mathrm{c}} \leqslant v_{\mathrm{esc}}$ and
$0 \leqslant v_{\mathrm{r}} \leqslant 2 v_{\mathrm{esc}}$.

Finally, we note that the product $f_{\vec{v}}(\vec{v}_{1},r)\,f_{\vec{v}}(\vec{v}_{2},r)$ is conserved
in the transformation $\mu \rightarrow - \mu$. Thanks to this symmetry, it is sufficient to
perform the integral over $\mu$ between 0 and $\mu_{0}$. 

The final expression for the relative velocity DF is therefore
\beq
F_{\rm r}(\vec{v}_{\rm r},r) = 4 \pi \kappa^{-1}(r) \int_{0}^{v_{\mathrm{esc}}} \!
{\rm d}v_{\rm c} \, v_{\rm c}^{2} \int_{0}^{\mu_{0}} \! \mathrm{d}\mu \,
f_{\vec{v}}(v_{1},r)\,f_{\vec{v}}(v_{2},r),
\eeq
If $f_{\vec{v}}(\vec{v},r)$ is determined self-consistently and normalized to 1 by construction,
then the normalization function $\kappa(r)=1$. However, when modifying the DF, typically to account
for the divergence discussed in \citesec{ssec:rmax}, one needs to compute $\kappa(r)$
for each value of $r$ and renormalize the relative velocity DF by hand.

One can also define the distribution function for the relative speed $v_{\rm r} = |\vec{v}_{\rm r}|$.
For an isotropic system, this reads
\beq
F_{\rm r}^{\mathrm{1D}}(v_{\rm r},r) = 4 \pi v_{\rm r}^{2} F_{\rm r}(\vec{v}_{\rm r},r).
\eeq
This allows one to readily compute the average of any observable $\mathcal{O}(v_{\rm r})$ via
\ben
\langle \mathcal{O} \rangle_{v_{\rm r}}(r) =
\kappa^{-1}(r) \int_{0}^{2 \, v_{\mathrm{esc}}(r)} \!
{\rm d}v_{\rm r} \, F_{\rm r}^{\mathrm{1D}}(v_{\rm r},r) \mathcal{O}(v_{\rm r}).
\een
\subsection{Anisotropic extensions}
\begin{figure*}[!t]
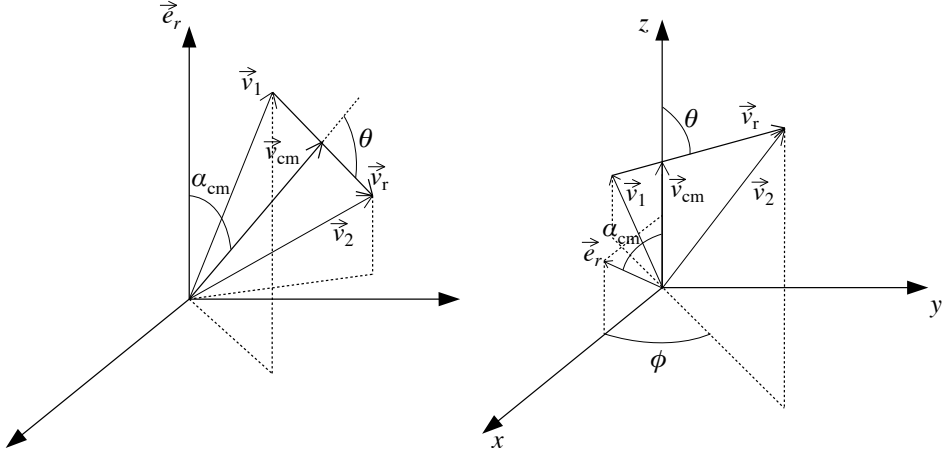

\centering
\includegraphics[width=0.4\linewidth]{{{fig_diagram1}}}
\includegraphics[width=0.4\linewidth]{{{fig_diagram2}}}
\caption{\small Coordinate systems for the derivation of the relative velocity DF for an
  anisotropic system. \textbf{Left panel:} Frame associated with $\vec{r}$, in which we define
  $\vec{v}_{\mathrm{c}}$. \textbf{Right panel:} Coordinate system associated with
  $\vec{v}_{\mathrm{c}}$.}
\label{fig:coordinate_systems}
\end{figure*}
For an anisotropic system, the DF no longer depends on the position and velocity only, but it also
depends on the modulus of the angular momentum, $L$: $f \equiv f(v,L,r)$. In this case, it is still
convenient to perform a change of variables from $\vec{v}_{1}$, $\vec{v}_{2}$ to
$\vec{v}_{\mathrm{c}}$, $\vec{v}_{\mathrm{r}}$, provided one uses the appropriate coordinate systems
to describe the quantities of interest.

The anisotropy of the system is characterized by the specific direction defined by the radial unit
vector $\vec{e}_{r}$. Physically, the outer integral on $\vec{v}_{\mathrm{c}}$ is equivalent to fixing
$v_{\mathrm{c}} = |\vec{v}_{\mathrm{c}}|$ and defining the angle
\ben
\alpha_{\mathrm{c}} \equiv (\vec{v}_{\mathrm{c}}, \vec{e}_{r})
\een
in the coordinate system associated with $\vec{e}_{r}$, which is illustrated in the left panel of
\citefig{fig:coordinate_systems}. Once $\vec{v}_{\mathrm{c}}$ is fixed, the system is invariant under
any rotation about $\vec{e}_{r}$, so the associated angular integral directly gives a factor
$2\pi$.

To define $\vec{v}_{\mathrm{r}}$, we then use the frame defined by $\vec{v}_{\mathrm{c}}$, illustrated
in the right panel of \citefig{fig:coordinate_systems}. There is no loss of generality by assuming
that $\vec{e}_{r}$ is orthogonal to $\vec{e}_{y}$. Therefore, in this frame, the vectors of interest
read
\begin{subequations}
\label{eq:anis_vectors}
  \ben
\vec{v}_{\mathrm{c}} & =& v_{\mathrm{c}}\ \vec{e}_{z}, \\
\vec{e}_{r} & =& \cos \alpha_{\mathrm{c}}\ \vec{e}_{z} + \sin \alpha_{\mathrm{c}}\ \vec{e}_{x}, \\
\vec{v}_{\mathrm{r}} & = & v_{\mathrm{r}}
\left( \cos \theta\ \vec{e}_{z} + \sin \theta \cos \phi\ \vec{e}_{x} 
+ \sin \theta \sin \phi\  \vec{e}_{y}  \right).
\een
\end{subequations}
In the frame fixed by $\vec{v}_{\mathrm{c}}$, the situation is the same as in the isotropic case,
and we can use the integral bounds derived in the previous section. The DF for the modulus of the
relative velocity is then given by 
\ben
F_{\rm r}^{\mathrm{1D}}(v_{\rm r},r) =  2 \, \pi\,\kappa^{-1}(r)\, v_{\rm r}^{2}
\int_{0}^{v_{\mathrm{esc}}} \! {\rm d}v_{\mathrm{c}} \, v_{\mathrm{c}}^{2}
\int_{-1}^{1} \! {\rm d}\mu_{\mathrm{c}}
\int_{0}^{2\pi} \! {\rm d}\phi \, \int_{-\mu_{0}}^{\mu_{0}} \! \mathrm{d}\mu \,
f_{\vec{v}}(v_{1},L_{1},r) f_{\vec{v}}(v_{2},L_{2},r)\,,\nn\\
\een
where $\mu_{\mathrm{c}} = \cos \alpha_{\mathrm{c}}$ and $\mu = \cos \theta	$. The velocity moduli
$v_{1}$ and $v_{2}$ are defined in \citeeq{eq:v1_v2}, and
\begin{subequations}
\ben
L_{1}^{2} & = &\left|\vec{r} \times
\left( \vec{v}_{\rm c} - \dfrac{\vec{v_{\rm r}}}{2} \right) \right|^{2}  \nn \\
& = & r^{2}
\left[ \dfrac{v_{\mathrm{r}}^{2}}{4} (1 - \mu^{2}) \sin^{2} \phi 
  + \left\lbrace - \dfrac{v_{\mathrm{r}}}{2} \mu_{\mathrm{c}} \sqrt{1-\mu^{2}} \cos \phi -
  \sqrt{1 - \mu_{\mathrm{c}}^{2}} (v_{\mathrm{c}} -\dfrac{v_{\mathrm{r}}}{2} \mu) \right\rbrace ^{2}
  \right] \,,\\
L_{2}^{2} & = &\left|\vec{r} \times \left( \vec{v}_{\rm c} + \dfrac{\vec{v_{\rm r}}}{2} \right) \right|^{2}  \nn \\
& = & r^{2} \left[ \dfrac{v_{\mathrm{r}}^{2}}{4} (1 - \mu^{2}) \sin^{2} \phi
  + \left\lbrace \dfrac{v_{\mathrm{r}}}{2} \mu_{\mathrm{c}} \sqrt{1-\mu^{2}} \cos \phi -
  \sqrt{1 - \mu_{\mathrm{c}}^{2}} (v_{\mathrm{c}}  +\dfrac{v_{\mathrm{r}}}{2} \mu) \right\rbrace ^{2}
  \right] \,.
\een
\end{subequations}

\bibliographystyle{JHEP}
\bibliography{biblio_Eddington_v1}

\end{document}